\documentclass[prx,aps,letterpaper,twocolumn,allowfontchageintitle=true,unpublished]{quantumarticle}

\usepackage{bbm}

\usepackage{dcolumn}
\usepackage{bm}
\usepackage{graphicx}
\usepackage{hyperref}
\usepackage{color}
\usepackage{version}
\usepackage{mathtools}
\graphicspath{{figures/}}
\usepackage{natbib}
\usepackage{multicol}
\usepackage{float}
\usepackage{wasysym}
\usepackage{tabularx}

\usepackage[table]{xcolor}
\usepackage{booktabs}
\usepackage{xcolor}

\begin{document}

\title{\centering \color{quantumviolet} Blocklet concatenation: Low-overhead fault-tolerant protocols for fusion-based quantum computation \newpage}
\author{\vspace{-5ex}}
\affiliation{Daniel Litinski @ PsiQuantum, Palo Alto}
\date{\vspace{-7ex}}
{\centering \maketitle}

\begin{abstract}

We introduce a construction for protocols for fault-tolerant quantum computing based on code concatenation and transversal gates. These protocols can be interpreted as families of quantum circuits of low-weight stabilizer measurements without strict locality constraints, effectively implementing concatenated codes. However, we primarily study these protocols in the context of photonic fusion-based quantum computing (FBQC), where they yield families of fusion networks with constant-sized resource states. Their high erasure thresholds relative to their resource-state cost establish them as promising candidates to replace surface codes in the context of FBQC. Examples include protocol families using 8-, 10- and 12-qubit resource states, with erasure thresholds of 13.8\%, 19.1\% and 11.5\%, and footprint-per-logical-qubit scaling as $\mathcal{O}(d)$, $\mathcal{O}(d^{1.46})$ and $\mathcal{O}(d^{0.58})$, respectively, where $d$ is the code distance. We also present techniques for performing logical operations, decoding, and implementing the protocols in photonic hardware. Although we focus on photonic FBQC, these ideas may also be of interest in other settings.

\end{abstract}

\section{Introduction}

\textbf{FBQC in a nutshell.}
Quantum computation is commonly described in terms of \textit{circuit-based quantum computation} (CBQC), where gates and measurements are applied to a static array of qubits.
In the context of quantum error correction, these operations are primarily used to repeatedly measure the stabilizers of a quantum error-correcting code. However, a photonic quantum computer based on linear optics features neither static qubits nor deterministic entangling gates. Instead, it is more convenient to use \textit{resource states} and \textit{fusions} as elementary building blocks. A \textit{fusion-based quantum computation} (FBQC)~\cite{Bartolucci2021} is performed by repeatedly generating copies of identical few-photon states (resource states) and performing entangling two-photon measurements (fusions) between pairs of photons from different resource states (though more general multi-qubit fusions are also possible). A prescription for executing a quantum computation using specific resource states and fusions (a \textit{fusion network}) is fault-tolerant, if it can tolerate imperfections in both resource states and fusions. Each fusion is a two-qubit Bell-basis measurement that generates two bits of information: a $Z \otimes Z$ and an $X \otimes X$ measurement outcome, where $X$ and $Z$ are Pauli operators. Each of these bits can be flipped or erased, either due to physical errors or due to the probabilistic nature of entangling operations. A fault-tolerant fusion network features redundancy in the form of parity checks that can be used to recover from flipped or erased bits. In other words, the fusion-outcome bits are encoded in an error-correcting code.

There is a simple prescription to convert any CBQC scheme for fault-tolerant quantum computation into an FBQC scheme, most conveniently expressed in terms of ZX calculus~\cite{Coecke2011, Backens2014, Coecke2017} as described in Ref.~\cite{Bombin2024}: Write down the quantum circuit of the computation (e.g., the syndrome readout circuit of a code), convert it into a ZX diagram, and partition the ZX diagram into identical chunks. The chunks can be viewed as resource states and the connections between different chunks as fusions. (More generally, one can also treat ZX spiders with $n$ legs as $n$-way multi-qubit fusions.) Any one ZX diagram can therefore give rise to many different fusion networks. One can choose to partition the diagram into a large number of small resource states or a smaller number of large resource states. Small resource states are cheaper, but since the fusion network contains a larger number of error-prone fusions, the error tolerance will be lower. Conversely, larger resource states are more expensive, but lead to higher error tolerance, i.e., a higher threshold error rate. In addition to straightforwardly converting stabilizer readout circuits of codes to fusion networks (which is also referred to as \textit{foliation}~\cite{Bolt2016}), one can directly generate a plethora of fusion networks of topological codes from geometric considerations, e.g., via fault-tolerant complexes~\cite{Bombin2023}. This includes not only many variations of 2D surface codes~\cite{Kitaev2003,Bravyi1998}, but also color codes~\cite{Bombin2006} and higher-dimensional codes.

\renewcommand{\arraystretch}{1.1}

\begin{table*}[t]
\centering
\begin{tabular}{llll}
Protocol family & Resource state size & Erasure threshold & Footprint per logical qubit \\
\hline
2D surface code (cubic~\cite{Bartolucci2021}) & 6 qubits & $12.0\%$ & $d^2$ resource states \\
2D surface code (cuboctahedral~\cite{Bombin2023}) & 8 qubits & $12.7\%$ & $\frac{3}{4}d^2$ resource states \\
$[4, 2, 2]$ blocklet & 8 qubits & $13.8\%$ & $\frac{1}{4}d$ resource states \\
$[5, 1, 3]$ blocklet & 10 qubits & $19.1\%$ & $\approx 0.47d^{1.46}$ resource states \\
$[6, 4, 2]$ blocklet & 12 qubits & $11.5\%$ & $\approx \frac{1}{6}d^{0.58}$ resource states \\
$[7, 1, 3]$ blocklet & 14 qubits & $19.2\%$ & $\approx 0.22d^{1.77}$ resource states \\
\end{tabular}
\caption{Erasure thresholds and footprint scaling behavior of a selection of fault-tolerant protocols discussed in this paper. Note that the footprint scaling of the blocklet protocols is based on the conjecture described in Appendix~\ref{app:distance}.}
\label{tab:results}
\end{table*}

\textbf{The surface code barrier.}
The search for good fusion networks is a search for favorable footprint-to-threshold trade-offs. When comparing two fusion networks with different thresholds, the higher-threshold fusion network is only better, if its footprint per logical qubit is not incommensurately high. Even when restricted to small resource states, the number of protocols that can be generated via fusion complexes or foliation of various qLDPC codes~\cite{Tillich2014, Panteleev2022, Bravyi2024} is extremely large. Despite the vast search space, it is remarkably challenging to find FBQC protocols that outperform the simplest topological protocols based on the 2D surface code, i.e., 3-dimensional fusion networks in which each fusion outcome contributes to at most two checks~\cite{Bombin2023}. This motivates the exploration of alternative prescriptions to generate fault-tolerant protocols.

In this paper, we introduce a new method for the construction of fault-tolerant protocols that is different from foliation or from constructions based on geometric considerations. Instead, it is based on concatenation and transversal gates. The study of concatenated quantum codes dates back to the early days of quantum computation~\cite{Knill1996, Aharonov1999, Aliferis2005}, and has received some recent attention in the context of theoretical results on constant-overhead quantum computation~\cite{Yamasaki2024, Yoshida2024, Gidney2025}. 
Several recent works have also studied constructions based on tensor networks~\cite{Ferris2014, Cao2022, Cao2024, Steinberg2025}. Here, we focus specifically on constructions of protocols (rather than codes) in the context of photonic FBQC. We refer to the presented method as \textit{blocklet concatenation}.

\textbf{Blocklet concatenation.} The set of transversal operations of any CSS code includes all operations that can be expressed as ZX diagrams without phases or Hadamards. This includes operations such as CNOT gates, ZZ measurements, XX measurements and GHZ-state projections. The syndrome readout circuit of an $n$-qubit CSS code is an $n$-qubit operation that can be written as a ZX diagram without phases or Hadamards. Therefore, the readout circuit of any CSS code is itself a transversal operation of that code.
In other words, a syndrome readout operation on $n$ logical qubits can be performed by applying $n$ copies of the physical $n$-qubit readout circuit, each acting transversally across the $n$ code blocks.
We can use this to repeatedly concatenate a CSS code with itself while only relying on syndrome readout circuits of the original CSS code. As a resource state, the syndrome readout circuit of an $n$-qubit code corresponds to a $2n$-qubit state that we refer to as a \textit{blocklet}. This state is a Bell pair in which both qubits are encoded in the $n$-qubit code. We can therefore construct families of FBQC protocols with increasing levels of concatenation (and therefore increasing code distances) that rely on identical blocklet resource states. 

We describe this construction in detail in Sec.~\ref{sec:construction} and find that it not only works for CSS codes, but also for some non-CSS codes such as the five-qubit code. It can also be used with codes that encode multiple logical qubits, in which case each concatenation step also increases the number of logical qubits. When repeatedly concatenating an $[n, k, d]$ code with $n$ physical qubits, $k$ logical qubits and a code distance of $d$, each concatenation step increases the footprint by a factor of $n$ and the number of logical qubits by a factor of $k$, whereas the size of the resource state remains constant. We also conjecture the code distance to scale with $\mathcal{O}(d^L)$, where $L$ is the number of concatenation steps. Therefore, the footprint can scale with the code distance substantially more favorably compared to the $\mathcal{O}(d^2)$ scaling of surface codes. It is also possible to mix different codes in one protocol.

\begin{figure*}[t!]
\centering
\includegraphics[width=0.99\linewidth]{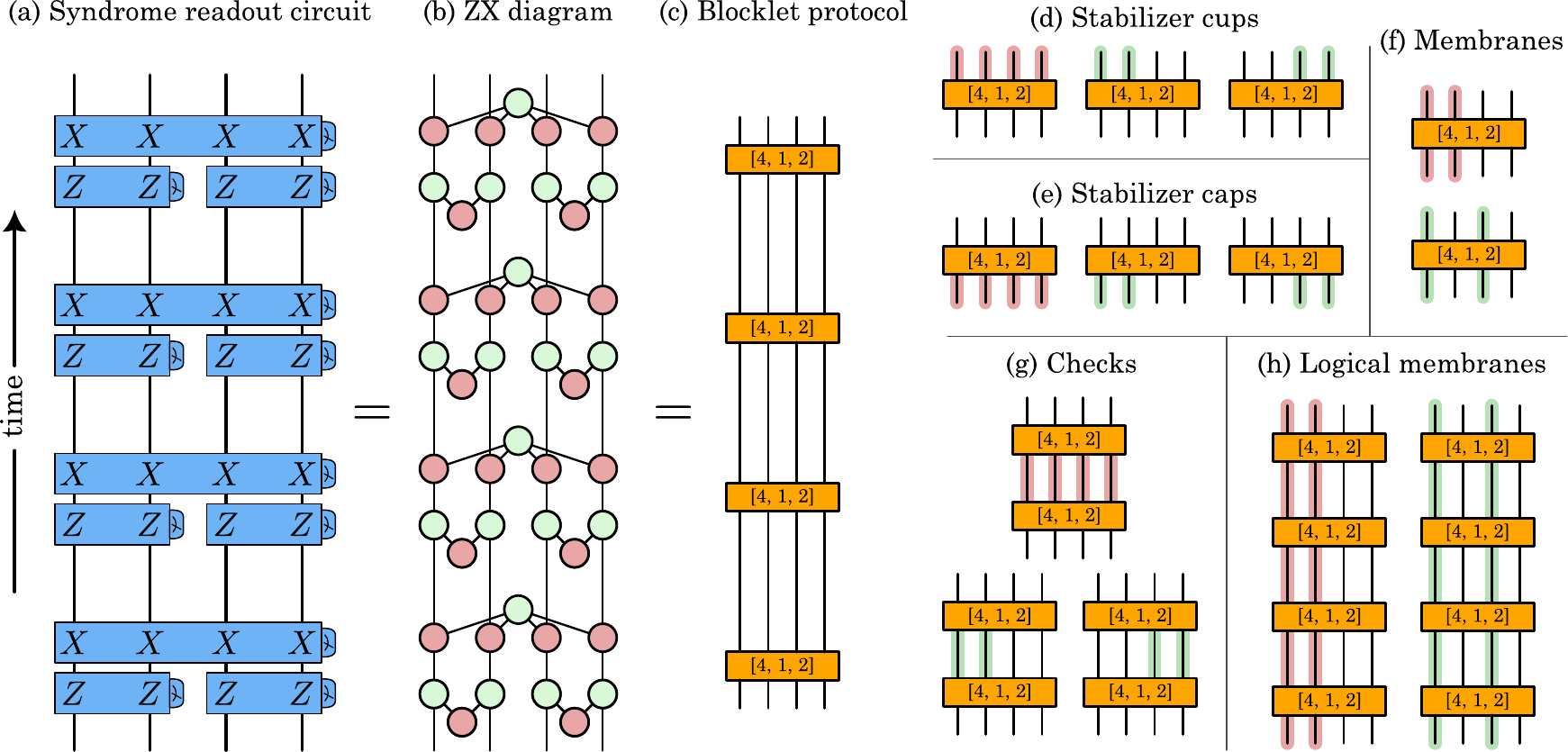}
\caption{Introduction to blocklets. A syndrome readout circuit (a) of a stabilizer code can be converted into a ZX diagram (b) which can be interpreted as a blocklet protocol (c), i.e., a fusion network in which each orange box is a blocklet resource state and each edge is a fusion. Blocklets have various Pauli webs (d-f) that combine to form checks (g) and logical membranes (h) in a fault-tolerant protocol.}
\label{fig:blockletintro}
\end{figure*}

\textbf{Better than surface codes?} Remarkably, many of these protocol families correspond to fault-tolerant fusion networks with high erasure thresholds. In photonic FBQC, the main source of errors is photon loss leading to erased fusion outcomes. Typically, fusion networks with higher erasure thresholds also rely on more expensive resource states. For example, Table~\ref{tab:results} highlights two surface-code fusion networks, one using a 6-qubit resource state with a 12\% erasure threshold, and another using an 8-qubit resource state with a 12.7\% erasure threshold. The blocklet protocols that we construct in this paper often feature more expensive resource states, but this is mitigated by both the better footprint scaling and a disproportionately higher erasure threshold. For example, the blocklet protocols highlighted in Table~\ref{tab:results} include the $[4, 2, 2]$ blocklet protocol with an 8-qubit resource state, a 13.8\% erasure threshold and $\mathcal{O}(d)$ footprint scaling, as well as the $[5, 1, 3]$ blocklet protocol with a 10-qubit resource state, a 19.1\% erasure threshold and a $\mathcal{O}(d^{1.46})$ footprint scaling. While the number of qubits in a resource state is not always a reliable predictor of preparation cost---since details of the physical primitives can significantly affect relative overhead---it remains a useful starting point for comparison~\cite{Bartolucci2025}, and we still expect these protocols to be competitive.

\textbf{Low-weight, but not LDPC.} Much of the search for low-overhead fault-tolerant protocols has focused on LDPC codes due to their low-weight stabilizer measurements. Lower-weight measurements are desirable because they involve fewer physical components and are thus less susceptible to noise. Our blocklet concatenation protocols also use only low-weight measurements, yet are not LDPC, as they feature a hierarchy of increasingly larger checks. Complementing recent results that highlight concatenation as a promising alternative to LDPC codes~\cite{Yamasaki2024, Yoshida2024, Gidney2025}, our results demonstrate that one can repeatedly concatenate an $[n,k,d]$ code while only relying on the low-weight measurements of the base code. More importantly, this approach yields high-threshold, low-footprint fault-tolerant protocols.

\textbf{Structure of the paper.} After describing the protocol construction in Sec.~\ref{sec:construction}, we discuss in Sec.~\ref{sec:logic} how to perform logical operations with these protocols. We show that blocklets are compatible with the active-volume architecture~\cite{Litinski2022}, implying that quantum computations can be compiled to and executed with logical operations in a manner similar to efficient surface-code architectures. In Sec.~\ref{sec:decoding}, we discuss decoding strategies for blocklet protocols. While erasure-only simulations can be performed with an efficient optimal decoder, the noise in a real device will also lead to Pauli errors, even if the Pauli error rate may be significantly lower than the erasure rate. The presence of Pauli errors necessitates an efficient decoding scheme for blocklets. As a proof of principle, we describe a simple \textit{hierarchical decoding} scheme that can decode a distance-$d$ \linebreak protocol in $\mathcal{O}(\log d)$ sequential steps of parallelizable decoding tasks, but it underperforms compared to the optimal decoder. For the $[5, 1, 3]$ protocol family, the erasure threshold obtained in this manner is 13\% below the optimal threshold, i.e., the crossing point between the two largest simulated protocols is at an erasure rate of 16.7\% instead of 19.1\%. However, the simulations demonstrate that blocklet protocols also have an acceptable performance under Pauli noise, as this crossing point is at a 1.7\% Pauli error rate. Future improvements may close the performance gap. Finally, we discuss in Sec.~\ref{sec:photonic} how to implement blocklet protocols in photonic hardware. The increased performance of blocklets compared to surface codes comes at the cost of non-local operations, as the fusion network is no longer a regular lattice. Still, we show that they can be implemented with interleaving modules~\cite{Bombin2021} that contain $\mathcal{O}(\log d)$ switchable fiber delays.

\begin{figure*}[t!]
\centering
\includegraphics[width=\linewidth]{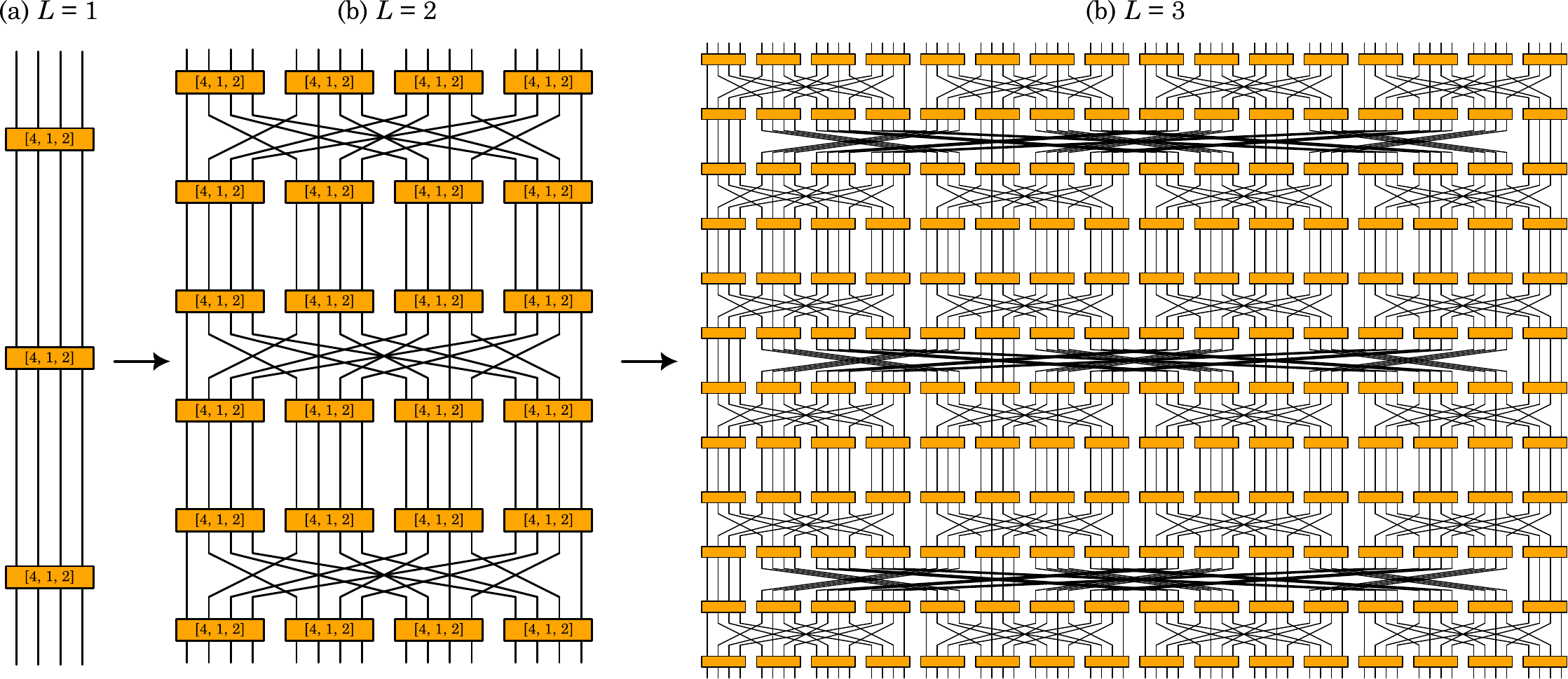}
\caption{Concatenated $[4, 1, 2]$ blocklet protocols at different concatenation levels $L$.}
\label{fig:concatenation}
\end{figure*}

\section{Protocol construction}
\label{sec:construction}

\textbf{Foliated protocols.} Consider the syndrome readout circuit of the $[4, 1, 2]$ code in Fig.~\ref{fig:blockletintro}a. Here, the $[4, 1, 2]$ code is a stabilizer code encoding one logical qubit in four physical qubits with stabilizers $X_1X_2X_3X_4$, $Z_1Z_2$ and $Z_3Z_4$. Its syndrome readout circuit corresponds to the logical identity gate of the code and is a sequence of stabilizer measurements. The circuit can be converted into the ZX diagram in Fig.~\ref{fig:blockletintro}b. The diagram features a repeating structure of 11 spiders. In the rest of the paper, we will implicitly treat our protocols as ZX diagrams, but instead of explicitly drawing pieces of ZX diagrams corresponding to a set of stabilizer measurements, we will represent them in a compressed manner using boxes as shown in Fig.~\ref{fig:blockletintro}c. We can interpret this ZX diagram as a fusion network by treating each box as a resource state and the connections between boxes as fusions, such that each edge incident on a box corresponds to one of its resource-state qubits. We refer to the $2n$-qubit state that is represented by the box of an $[n, k, d]$ syndrome readout circuit as an $[n, k, d]$ blocklet. The state represented by each box in Fig.~\ref{fig:blockletintro} has 8 qubits, which we will also refer to as \textit{ports}, i.e., four top ports and four bottom ports. The state has 8 stabilizers. In the language of ZX diagrams, each stabilizer is equivalent to a  Pauli web~\cite{Bombin2024}. We will refer to Pauli webs that are entirely supported on the top ports as \textit{stabilizer cups} (or simply \textit{cups}), to those supported entirely on the bottom ports as \textit{stabilizer caps} (or simply \textit{caps}) and to those supported on both bottom and top ports as \textit{membranes}. The Pauli webs of the $[4, 1, 2]$ blocklet are shown in Fig.~\ref{fig:blockletintro}d-f, with $X$ and $Z$ stabilizers highlighted in green and red, respectively. The stabilizers reveal that the state is a Bell pair with both qubits encoded in a $[4, 1, 2]$ code. In other words, blocklets are the same as encoded Bell pairs.

In the identity-gate protocol in Fig.~\ref{fig:blockletintro}c, blocklets are chained together to form a fault-tolerant protocol, a process which is also referred to as foliation. Whenever two blocklets are fused, their cups and caps combine to form checks, as shown in Fig.~\ref{fig:blockletintro}g. These are the parity checks of the fault-tolerant protocol. Each edge in the protocol corresponds to two classical bits, i.e., the $XX$ and $ZZ$ fusion outcomes of the corresponding fusion. The Pauli webs highlight the bits participating in the parity check---red for $ZZ$ outcomes and green for $XX$ outcomes. The logical information that is preserved by the protocol is described by the Pauli webs shown in Fig.~\ref{fig:blockletintro}h. These are the logical $X$ and $Z$ operators of the code (which we also call \textit{logical membranes}), and they are formed by chaining together the membrane Pauli webs of the different blocklets.

\begin{figure}[b!]
\centering
\includegraphics[width=\linewidth]{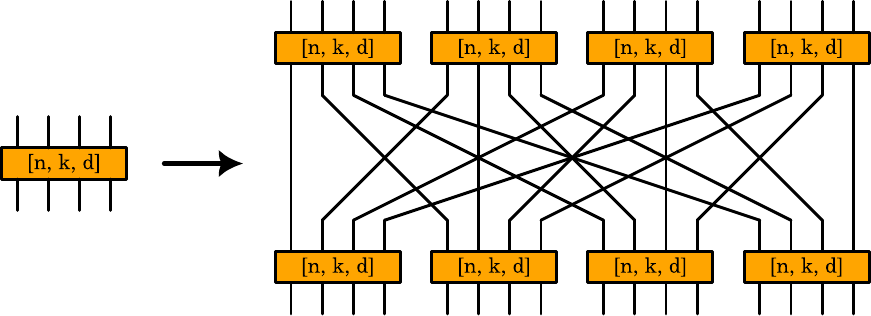}
\caption{Blocklet concatenation prescription.}
\label{fig:prescription}
\end{figure}

\begin{figure*}[t!]
\centering
\includegraphics[width=\linewidth]{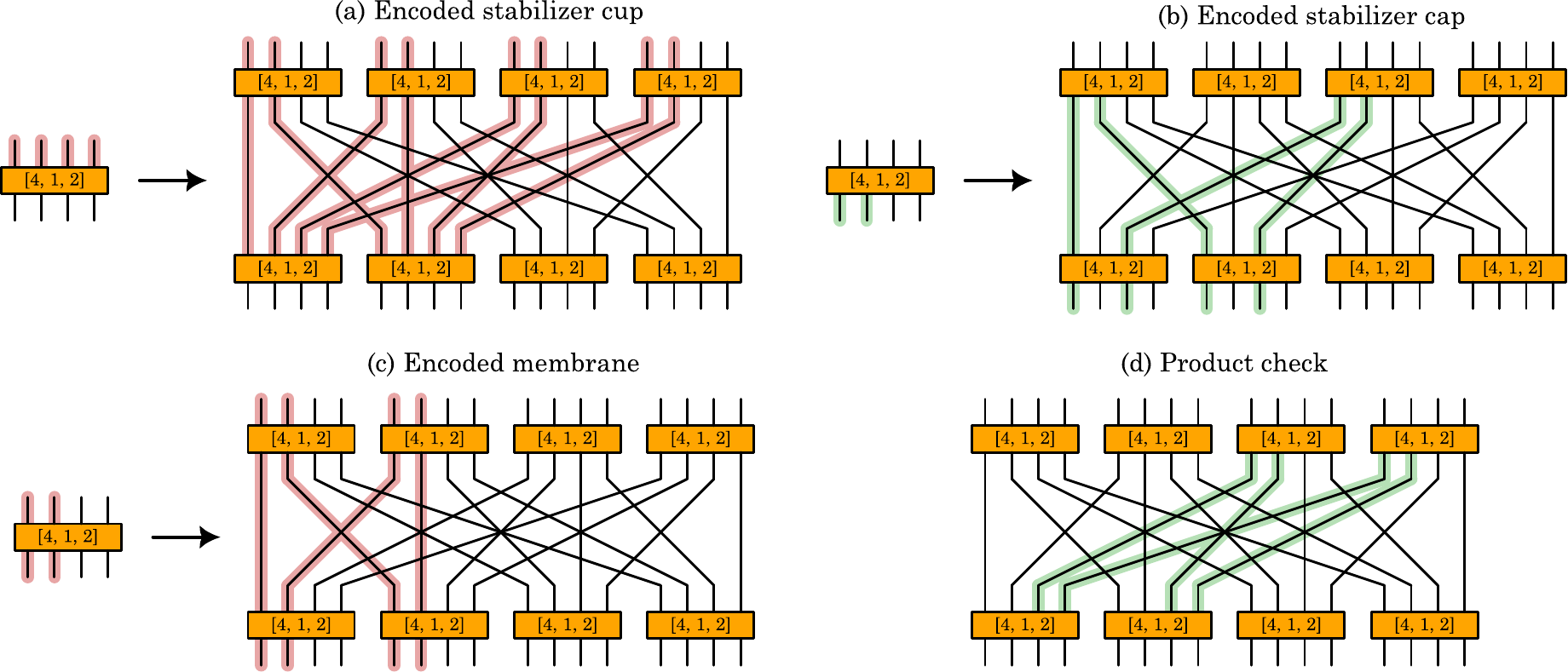}
\caption{Encoded Pauli webs are formed from combinations of Pauli webs of blocklets in a concatenation block.}
\label{fig:encodedwebs}
\end{figure*}

\textbf{Concatenation procedure.} Suppose we wanted to execute a transversal CNOT gate between two logical qubits encoded in a $[4, 1, 2]$ code. This would correspond to executing four physical CNOT gates. The $i$-th CNOT gate would act on the $i$-th physical qubits of the first and second logical qubit. If we want to transversally execute a four-qubit gate instead of a two-qubit gate, the prescription is similar: We execute four physical four-qubit gates such that the $i$-th physical gate acts on the $i$-th physical qubits of the four logical qubits. CNOT gates and single-qubit Pauli measurements are transversal gates of every CSS code. The syndrome-readout operation of every CSS code can be executed using only CNOT gates and single-qubit Pauli measurements. Therefore, the syndrome readout circuit of an $n$-qubit code is an $n$-qubit transversal operation of any CSS code. In particular, the 4-qubit operation corresponding to the $[4, 1, 2]$ syndrome readout circuit is a transversal operation of the $[4, 1, 2]$ code. Executing this operation on four $[4, 1, 2]$-encoded logical qubits is equivalent to effectively concatenating the codes to encode a single logical qubit in a $[16, 1, 4]$ code. When using this procedure with $[n, k, d]$ codes, this not only increases the code distance, but also increases the number of logical qubits by a factor of $k$.

The ZX diagram corresponding to the transversal execution of a $[4, 1, 2]$ syndrome readout circuit on four $[4, 1, 2]$-encoded logical qubits is shown in Fig.~\ref{fig:prescription}. We can use this to construct a simple diagrammatic prescription to transform fault-tolerant protocols into larger protocols with increased error suppression. In a protocol based on $[n, k, d]$ blocklets, we replace each blocklet with a group of $2n$ such blocklets split into two layers. We connect the $i$-th top port of the $j$-th blocklet in the bottom layer to the $j$-th bottom port of the $i$-th blocklet in the top layer. Each edge in the original protocol is replaced with a set of $n$ transversal edges between a pair of blocklets in the new protocol after the transformation. In Fig.~\ref{fig:concatenation} this prescription is used to repeatedly concatenate $[4, 1, 2]$ blocklets, gradually increasing the concatenation level $L$. We start with a foliated protocol in Fig.~\ref{fig:concatenation}a, which we refer to as the $L=1$ protocol. By applying the concatenation prescription of Fig.~\ref{fig:prescription}, we obtain the $L=2$ protocol in Fig.~\ref{fig:concatenation}b. If we apply this prescription again, we obtain the $L=3$ protocol in Fig.~\ref{fig:concatenation}c. We can repeatedly apply the concatenation prescription to obtain a family of protocols, where we refer to each protocol in the family as a $[4, 1, 2]^L$ protocol. Importantly, the elementary building blocks of the protocols remain identical, i.e., they can always be partitioned into $[4, 1, 2]$ blocklet resource states. What changes is the connectivity of the fusion network, which becomes increasingly non-local.

\textbf{Encoded Pauli webs.} While Fig.~\ref{fig:concatenation} describes the fault-tolerant protocol in terms of its operations, in practice, we also need a description of the checks and logical operators of the protocol. For $L=1$, the checks and logical operators of the protocol are those of the foliated protocol as described in Fig.~\ref{fig:blockletintro}. The checks of the foliated protocol are each obtained by combining one cup and one cap from adjacent blocklets. We will refer to such checks as \textit{fundamental checks} to distinguish them from other checks (\textit{product checks}) that are introduced later. When applying the concatenation prescription, the Pauli web of each blocklet is replaced with an \textit{encoded Pauli web}, thereby replacing each check with an \textit{encoded check}.

\begin{figure*}[t!]
\centering
\includegraphics[width=0.9\linewidth]{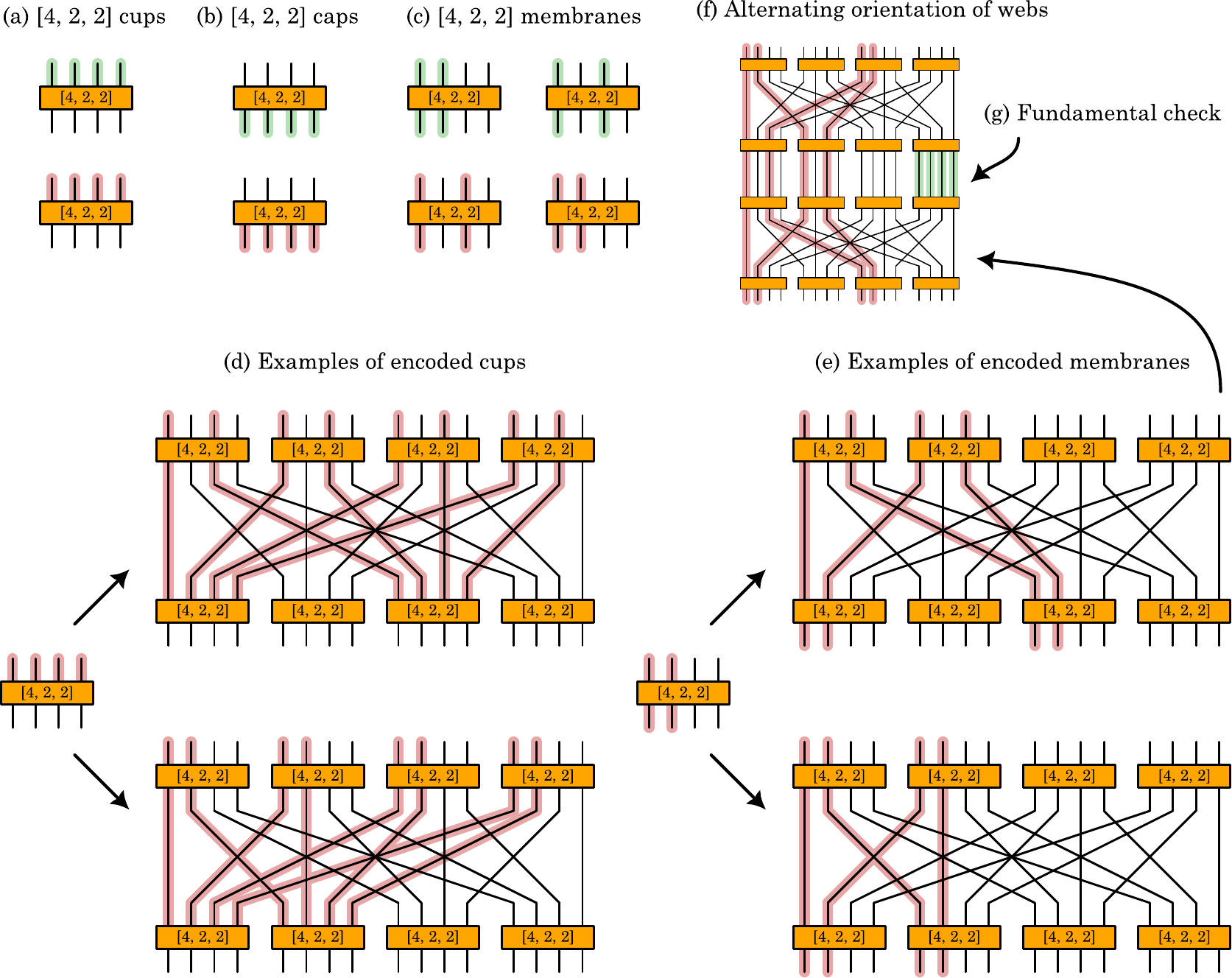}
\caption{In blocklet protocols with $k>1$, each encoded web has $k$ different variants, as shown for a $[4, 2, 2]$ code.}
\label{fig:422webs}
\end{figure*}

Consider only the $Z$-type Pauli webs of a blocklet. For every $Z$ stabilizer of the underlying code, there is one cup and one cap, and for every logical operator, there is one membrane. First, consider the case of $k=1$ such that there is only one $Z$ membrane. When replacing a blocklet via the concatenation prescription of Fig.~\ref{fig:prescription}, each cup is replaced with an \textit{encoded cup}. The encoded cup is formed from cups in the bottom layer and membranes in the top layer. If the logical $Z$ operator of the code is $Z_1^{j_1}Z_2^{j_2}\cdots Z_n^{j_n}$ (i.e., $j_i \in \{0, 1\}$ describe the support of the logical operator), then the $i$-th blocklet of the bottom layer contributes its cup to the encoded Pauli web only if $j_i = 1$. Similarly, if the stabilizer corresponding to the cup is $Z_1^{k_1}Z_2^{k_2}\cdots Z_n^{k_n}$, then the $i$-th blocklet of the top layer contributes its membrane to the encoded Pauli web only if $k_i = 1$. An example of an encoded cup of the $[4, 1, 2]$ code corresponding to the $Z_1Z_2Z_3Z_4$ stabilizer is shown in Fig.~\ref{fig:encodedwebs}a. Since the logical $Z$ operator is $Z_1Z_2$, the bottom layer contributes two cups from blocklets 1 and 2, and the top layer contributes four membranes from blocklets 1, 2, 3 and 4. The same procedure also applies to the cups in the $X$ basis.

Similarly, each cap turns into an encoded cap that is formed from caps in the top layer and membranes in the bottom layer. The support of the logical operator determines which blocklets contribute their caps, and the support of the stabilizer determines which blocklets contribute their membranes. The example in Fig.~\ref{fig:encodedwebs}b shows an encoded cap corresponding to the $X_1X_2$ stabilizer of the $[4, 1, 2]$ code. Since the logical $X$ operator is $X_1X_3$, the top layer contributes two caps from blocklets 1 and 3, and the bottom layer contributes two membranes from blocklets 1 and 2.

Finally, each membrane is replaced with an encoded membrane that is formed from membranes in the top and bottom layers. The encoded $Z$ membrane of the $[4, 1, 2]$ code is shown in Fig.~\ref{fig:encodedwebs}c.

The concatenation procedure transforms checks by replacing Pauli webs with their encoded versions. An example is shown in the appendix in Fig.~\ref{fig:checkconcatenation}. Here, the cups and caps of each fundamental check in $L=1$ are replaced with encoded cups and caps in $L=2$, and subsequently their constituent cups, caps and membranes are replaced with encoded cups, caps and membranes in $L=3$. Similarly, the constituent membranes of a logical membrane are replaced with their encoded versions in every concatenation step.

For codes with $k>1$, the procedure generates additional encoded cups, caps and membranes. There is one additional cup and cap for every additional logical operator of the code, generated via the same prescription as above, but using different logical operators and membranes. For example, consider the $[4, 2, 2]$ code. It has two stabilizers $X_1X_2X_3X_4$ and $Z_1Z_2Z_3Z_4$. The logical operators of the first logical qubit are $X_1X_2$ and $Z_1Z_3$ and the logical operators of the second logical qubit are $X_1X_3$ and $Z_1Z_2$. The $[4, 2, 2]$ blocklet is an 8-qubit state whose Pauli webs are shown in Fig.~\ref{fig:422webs}a-c. When generating encoded cups after concatenation, each cup turns into $k=2$ encoded cups, such as those shown in Fig.~\ref{fig:422webs}d. Similarly, each membrane turns into $k=2$ encoded membranes as shown in Fig.~\ref{fig:422webs}e. Therefore, checks are not only transformed with each concatenation steps, but the number of these checks is also multiplied by a factor of $k$. Similarly, the number of logical operators is multiplied by a factor of $k$ due to the additional encoded membranes. Note that when chaining together encoded membranes of the $[4, 2, 2]$ code or other codes with $k > 1$, every second encoded membrane is flipped vertically as shown in Fig.~\ref{fig:422webs}f.

\textbf{Additional checks.} The concatenation procedure not only transforms existing checks, but also introduces new ones. First, the concatenation procedure generates fundamental checks where blocklet pairs are fused transversally, i.e., one set of fundamental checks for each edge in the protocol prior to concatenation. As a reminder, fundamental checks are those that are formed from \textit{one} cup and \textit{one} cap. An example of such a fundamental check is shown in Fig.~\ref{fig:422webs}g. Furthermore, the concatenation procedure generates additional checks from the combination of cups from the bottom layer of blocklets and caps from the top layer. There is one such check for each pairing of stabilizers. We refer to these checks as \textit{product checks}. For example, a $Z$ product check that is generated from caps corresponding to a $Z$ stabilizer $Z_1^{j_1}Z_2^{j_2}\cdots Z_n^{j_n}$ and cups corresponding to a $Z$ stabilizer $Z_1^{k_1}Z_2^{k_2}\cdots Z_n^{k_n}$ is assembled the same way as other encoded webs: The $i$-th blocklet of the bottom layer contributes its cup if $j_i = 1$, and the $i$-th blocklet of the top layer contributes its cap if $k_i = 1$. An example for $[4, 1, 2]$ blocklets is shown in Fig.~\ref{fig:encodedwebs}d, where a product check is generated from cups of the $X_3X_4$ stabilizer and caps of the $X_1X_2$ stabilizer. A code with $m_X$ $X$ stabilizers and $m_Z$ $Z$ stabilizers will generate ${m_X}^2 + {m_Z}^2$ product checks for each blocklet after concatenation. In subsequent concatenation steps, these product checks will transform via encoded cups and caps the same way as fundamental checks.

In summary, the concatenation procedure generates a hierarchy of checks. There are small fundamental and product checks that are generated in the last concatenation step (\textit{level-1 checks}), and increasingly larger checks from previous concatenation steps. 
Examples of fundamental and product checks of various levels in an $L=3$ protocol are shown in Fig.~\ref{fig:checktypes}.
This is in stark contrast to protocols based on topological codes such as surface codes, where check sizes remain constant and have no hierarchical structure. When constructing a family of protocols by repeatedly concatenating $[n, k, d]$ codes, we will refer to members of this protocol family as $[n,k,d]^L$ protocols. An $[n, k, d]^L$ protocol has a footprint of $n^L$ qubits (or $n^{L-1}$ blocklet resource states) and encodes $k^L$ logical qubits.

\textbf{On the code distance.}
While the code distance of the concatenated code is $d^L$, determining the distance of the concatenated blocklet protocol is less trivial. We conjecture that the distance an $[n,k,d]^L$ protocol is $c \cdot d^L$, where $c$ is a constant that depends on the base code, e.g., $c=1$ for $[4, 2, 2]$ and $[6, 4, 2]$ codes, but $c=7/9$ for the $[7, 1, 3]$ code. As described in more detail in Appendix \ref{app:distance}, the first concatenation step encodes the fusions in the concatenation block of Fig.~\ref{fig:prescription} in a \textit{product code} that has a code distance $d_{\rm prod} \leq d^2$, which introduces error strings that may have a weight lower than $d^2$. Subsequent concatenation steps will increase the weight of these error strings by factors of $d$. We conjecture these to be the lowest-weight error strings, which is consistent with the subthreshold behavior observed in simulations. This implies that the code distance of an $[n,k,d]^L$ protocol is $d_p = c \cdot d^L$ with $c = d_{\rm prod}/d^2$, and that the footprint per logical qubit scales as $\mathcal{O}({\left(d_p/c\right)}^{\log(n/k)/\log(d)})$. In the language of FBQC, one may also view the concatenation procedure as a fault-tolerant way of adding \textit{local encoding} to the blocklet resource states.

\subsection{[5, 1, 3] blocklets}

\begin{figure}[t!]
\centering
\includegraphics[width=\linewidth]{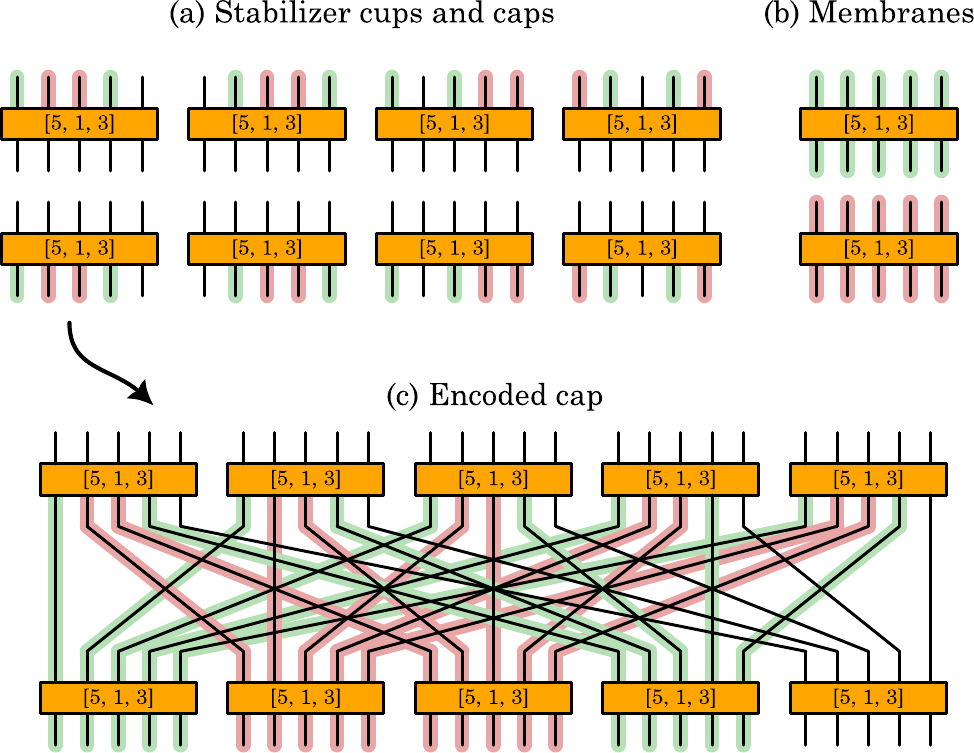}
\caption{Pauli webs of the $[5, 1, 3]$ blocklet and example of an encoded cap.}
\label{fig:513webs}
\end{figure}

The prescription described so far works for blocklets based on any CSS code. Surprisingly, the prescription also works for some non-CSS codes. One non-CSS code of interest is the $[5, 1, 3]$ code~\cite{Bennett1996,Laflamme1996}, as it is the smallest distance-3 code, and therefore will have better scaling properties under concatenation compared to the smallest distance-3 CSS code, which is the $[7, 1, 3]$ code. The stabilizers of the $[5, 1, 3]$ code are $X_1Z_2Z_3X_4$ and cyclic permutations thereof. The logical operators are $X_1X_2X_3X_4X_5$ and $Z_1Z_2Z_3Z_4Z_5$, although there are also lower-weight representatives such as $Z_1X_2Z_3$ and $Y_1Z_2Y_3$. A corresponding set of Pauli webs is shown in Fig.~\ref{fig:513webs}a-b.

We can generate a foliated protocol of $[5, 1, 3]$ blocklets and apply the concatenation procedure of Fig.~\ref{fig:prescription}, but this is only meaningful, if we also have a corresponding procedure to generate encoded Pauli webs and product checks. Encoded cups can be generated for every $X_{i_1}Z_{i_2}Z_{i_3}X_{i_4}$ stabilizer by combining the corresponding cups of all blocklets in the bottom layer with the $X$ membranes of blocklets $i_1$ and $i_4$ in the top layer, and the $Z$ membranes of blocklets $i_2$ and $i_3$ in the top layer. The analogous procedure for caps uses caps in the top layer and membranes in the bottom layer, as shown for the $X_1Z_2Z_3X_4$ cap in Fig.~\ref{fig:513webs}c. Encoded membranes are formed by combining the corresponding membranes of all 10 blocklets.

The nontrivial aspect are the product checks. The prescription for CSS codes combines $Z$ cups with $Z$ caps and $X$ cups with $X$ caps, but the cups and caps of the $[5, 1, 3]$ blocklet are mixed, making them incompatible with the CSS prescription. We can still find a set of 8 generating product checks from the combination of cups and caps that rely on the cyclic permutation symmetry of the $[5, 1, 3]$ code. One set of checks is shown in Fig.~\ref{fig:513productchecks}. This is not a general procedure that will work for any non-CSS code, but rather a special construction for $[5, 1, 3]$ codes. As mentioned in Appendix \ref{app:distance}, we find that the corresponding product code has a code distance of $d_{\rm prod} = 5$, implying a prefactor of $c = 5/9$ for the distance scaling.

\begin{figure}[t!]
\centering
\includegraphics[width=\linewidth]{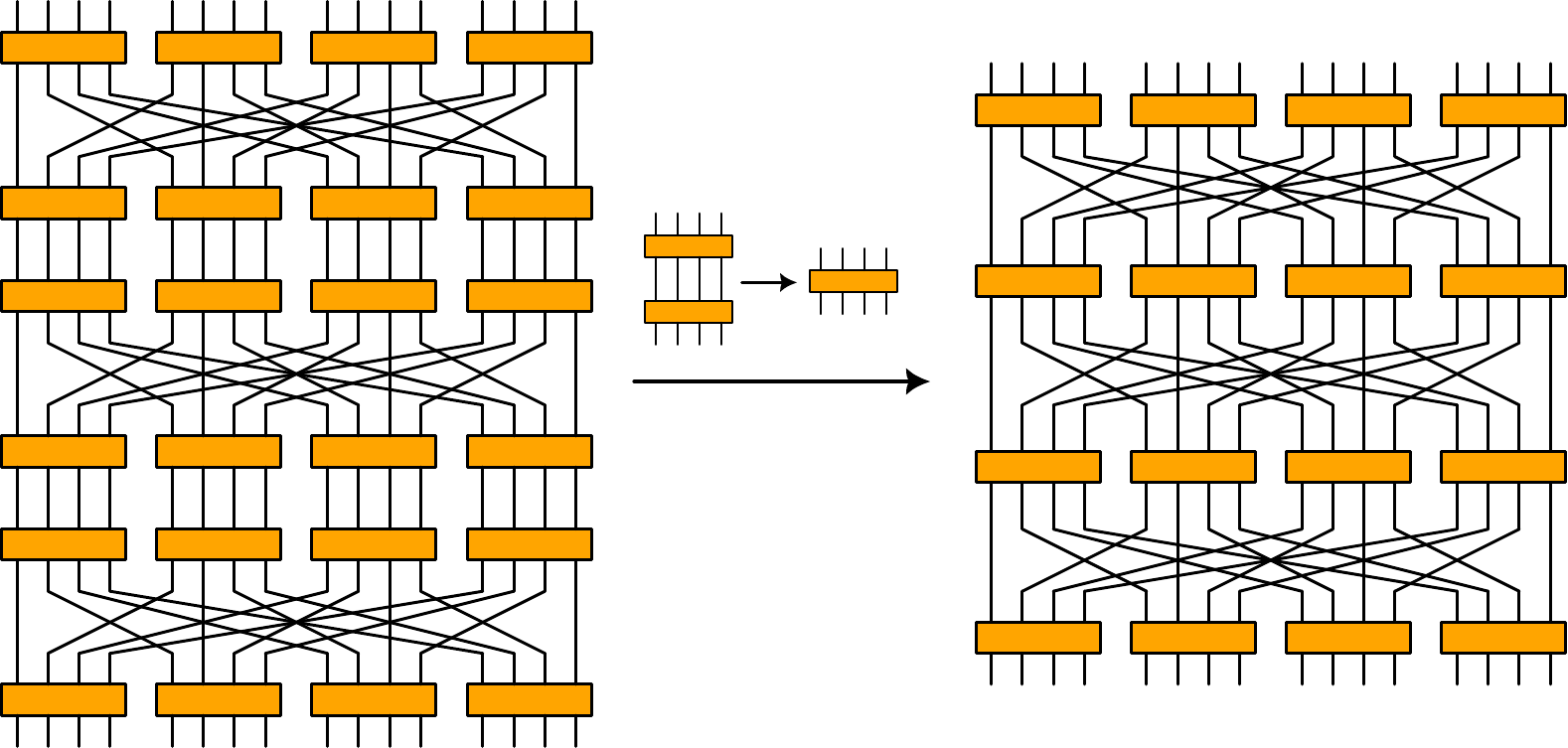}
\caption{Collapsing blocklet pairs after the final level of concatenation.}
\label{fig:collapse}
\end{figure}

\begin{figure*}[t!]
\centering
\includegraphics[width=0.99\linewidth]{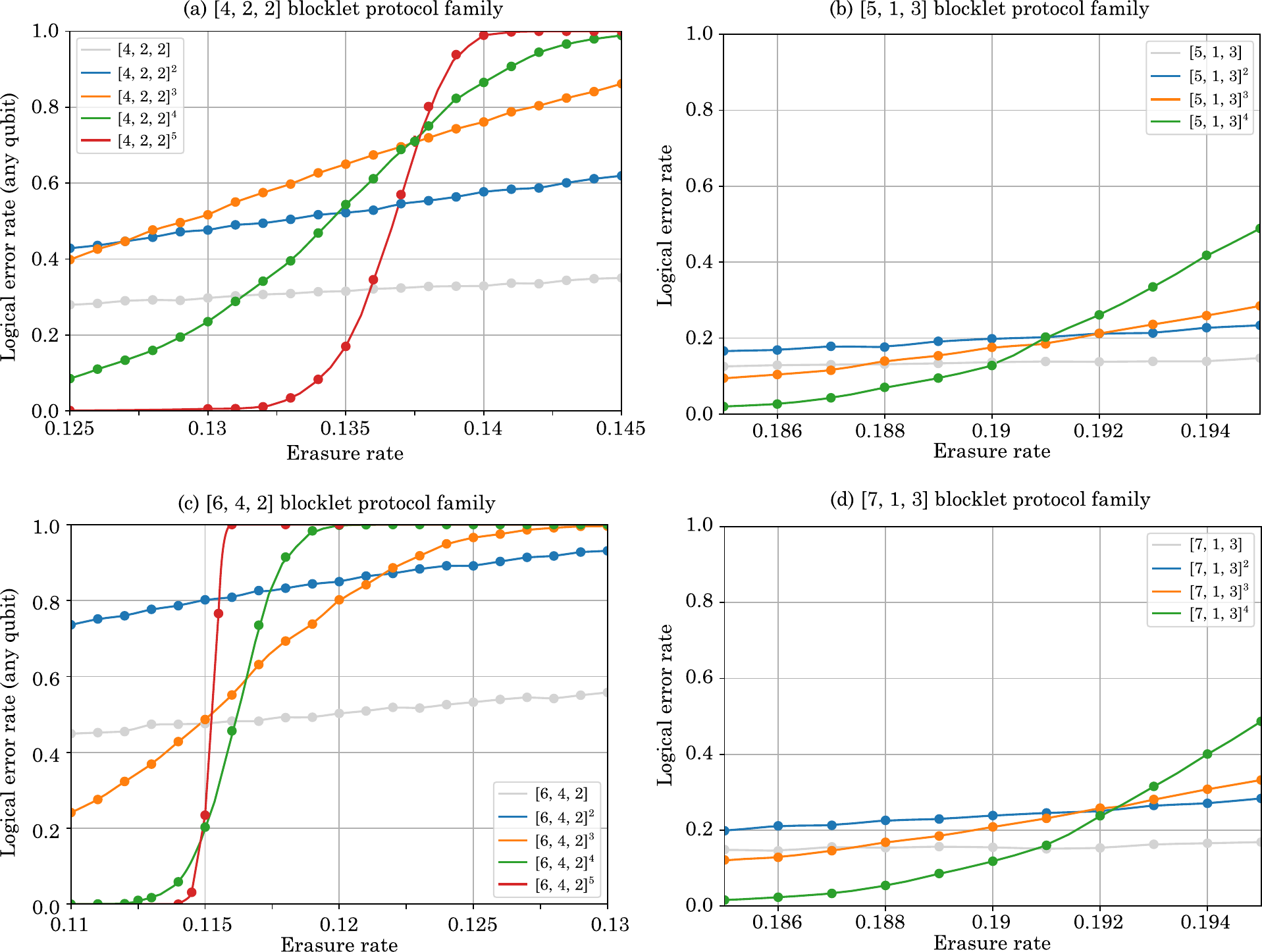}
\caption{Monte Carlo simulation results for various concatenated blocklet protocol families under erasure noise using an optimal erasure decoder.}
\label{fig:erasureplots}
\end{figure*}

\subsection{Erasure simulations}

We simulate the performance of the protocol families under an erasure noise model with independently and identically distributed erasures of $XX$ and $ZZ$ fusion outcomes. We consider families of protocols that are generated from foliated $L=1$ protocols with 8 layers of foliation---chosen arbitrarily to have a sufficiently large protocol to mitigate finite-size effects. As is common in such simulations, we also use periodic timelike boundary conditions to have well-defined logical operators for decoding and to further reduce finite-size effects. We generate concatenated protocols by repeatedly applying the concatenation prescription. After the final concatenation step, the protocol will consist of many transversally fused blocklet pairs. We halve the number of resource states and fusions in the protocol by collapsing all blocklet pairs into single blocklets, as shown in Fig.~\ref{fig:collapse}. Every concatenation step doubles the number of layers and increases the size of each layer by a factor of $n$.
Therefore, after collapsing the blocklet pairs, any $[n, k, d]^L$ protocol in our simulations will consist of $2^{L+1} \cdot n^L$ fusions and $2^{L+1} \cdot n^{L-1}$ resource states. We generate samples of erasure configurations in a Monte Carlo simulation and decode the erasures using an optimal erasure decoder based on Gaussian elimination. We report the logical error rate. For protocols encoding multiple logical qubits, an error on \textit{any} of the logical qubits counts as a logical error, e.g., for the $[6, 4, 2]^5$ protocols, this is an error on any of the $k^5= 1024$ logical qubits. 

A collection of simulation results is shown in Fig.~\ref{fig:erasureplots}. These are protocols based on the $[4, 2, 2]$ code, the $[5, 1, 3]$ code, the $[7, 1, 3]$ code (i.e., the Steane code) and the $[6, 4, 2]$ code with stabilizers $X^{\otimes 6}$ and $Z^{\otimes 6}$. We observe that the crossing point between logical error rate lines tends to converge with increasing levels of concatenation. We report the thresholds in Table~\ref{tab:results} as the physical error rates at which the logical error rates of the two largest simulated protocols intersect. For $[4, 2, 2]$ and $[6, 4, 2]$ blocklet protocols, these are the $L=4$ ($d_p=16$) and $L=5$ ($d_p=32$) protocols, whereas for $[5, 1, 3]$ and $[7, 1, 3]$ blocklet protocols, these are the $L=3$ ($d_p=15$ or $d_p=21$) and $L=4$ ($d_p=45$ or $d_p=63$) protocols. We also observe that the logical error rates indeed scale no worse than $\mathcal{O}(p^{d_p})$ below threshold, where $p$ is the erasure rate.

\textbf{Asymmetric codes.} The codes highlighted in Fig.~\ref{fig:erasureplots} feature equal protection against $X$ and $Z$ errors. If we build protocol families using codes that are not self-dual with respect to their stabilizers (or feature other symmetries like the $[5, 1, 3]$ code), we can obtain protocols that have different thresholds with respect to $XX$ and $ZZ$ erasures. Codes of interest in this category include tree codes, i.e., codes that are obtained from concatenating $X$-type and $Z$-type repetition codes, as their resource states are particularly cheap to prepare due to their ZX diagram exhibiting a tree structure. This includes the previously discussed $[4, 1, 2]$ code, which is a $(2,2)$-tree code, and the $[9, 1, 3]$ Shor code, which is a $(3,3)$-tree code. For the $[4, 1, 2]$ code, we find that it has a 21\% $XX$ erasure threshold, but only a 6.7\% $ZZ$ erasure threshold, as shown in Fig.~\ref{fig:treeblocklets}a-b. Similarly, the $[9, 1, 3]$ blocklet protocols have an $XX$ erasure threshold close to 38\% , but a $ZZ$ erasure threshold of only 6\% , as shown in Fig.~\ref{fig:treeblocklets}c-d. Note that there are several codes that can be described as $[9, 1, 3]$ codes. This particular tree code is a code with stabilizer generators $X_1X_2$, $X_2X_3$, $X_4X_5$, $X_5X_6$, $X_7X_8$, $X_8X_9$, $Z_1Z_2Z_3Z_4Z_5Z_6$, and $Z_4Z_5Z_6Z_7Z_8Z_9$. Blocklet protocols based on other $[9, 1, 3]$ codes, such as distance-3 surface codes, will perform differently.

While blocklet protocols based on tree codes may have cheap resource states, they cannot exceed a $\mathcal{O}({d_p}^2)$ footprint scaling due to the code parameters of tree codes. However, blocklet protocols with better footprint scaling may still have resource states that are trees. For example, the blocklet based on the $[6, 2, 2]$ code with stabilizer generators $X_1X_2$, $X_3X_4$, $X_5X_6$ and $Z_1Z_2Z_3Z_4Z_5Z_6$ is indeed a tree state, as shown in Fig.~\ref{fig:treeblocklets}g, even though the $[6, 2, 2]$ code is not a tree code. This particular protocol has a 30\% $XX$ erasure threshold and a 4.6\% $ZZ$ erasure threshold.

\begin{figure}[t!]
\centering
\includegraphics[width=\linewidth]{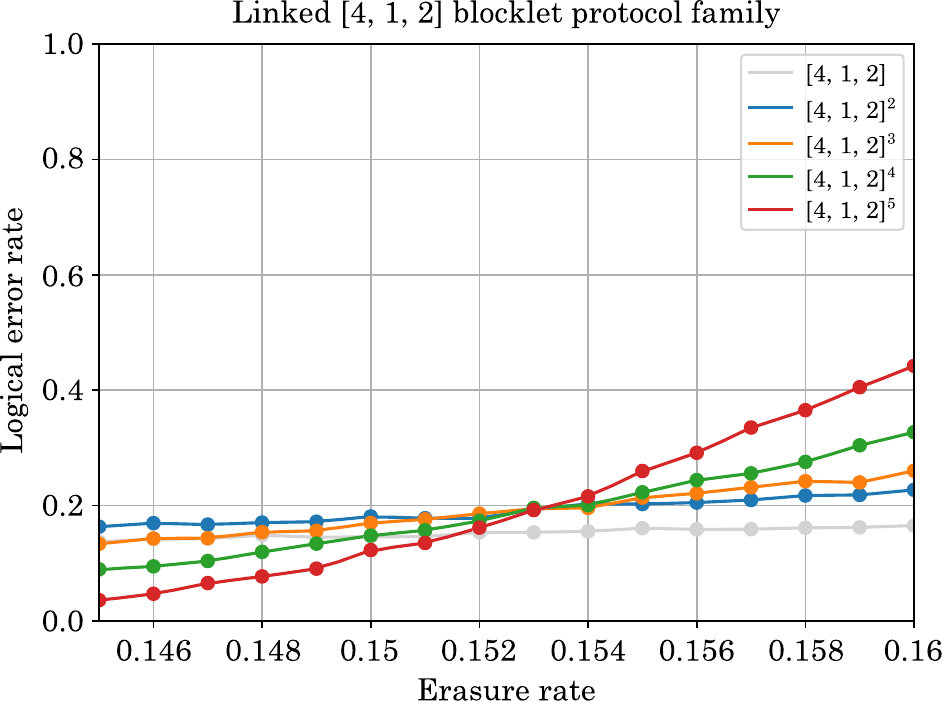}
\caption{Performance of the linked $[4, 1, 2]$ blocklet protocol.}
\label{fig:linked412}
\end{figure}

\textbf{Linking.} In the context of FBQC, one can make use of protocols with asymmetric $XX$ and $ZZ$ erasure thresholds by using an asymmetric local encoding~\cite{Bartolucci2021} that balances out the asymmetry. Another approach is to bias towards one type of erasure using dynamic bias arrangement~\cite{Bombin2023a}. One can also apply single-qubit Clifford gates to some of the qubits of the resource state prior to fusions to turn $XX$ fusion outcomes into $YY$ fusion outcomes. With uncorrelated $XX$ and $ZZ$ erasures, this increases the weight of the $X$ checks, as each $YY$ fusion outcome needs to be assembled using both the $XX$ and $ZZ$ outcome. But this also means that $ZZ$ outcomes now participate in more checks, increasing their error tolerance. We refer to this procedure as \textit{linking}, since it links the previously disconnected $XX$ and $ZZ$ parity check matrices. The choice of which fusions to apply Clifford gates to and which to leave unchanged affects the performance. For $[4, 1, 2]$ blocklets, we find that if we apply this transformation to all fusions except for every fourth fusion, we obtain a protocol family with a 15.3\% erasure threshold under uncorrelated $XX$ and $ZZ$ erasure, as shown in Fig.~\ref{fig:linked412}. However, in practice, $XX$ and $ZZ$ erasures feature correlation which may substantially diminish the apparent benefits of linking.

\begin{figure}[t!]
\centering
\includegraphics[width=0.97\linewidth]{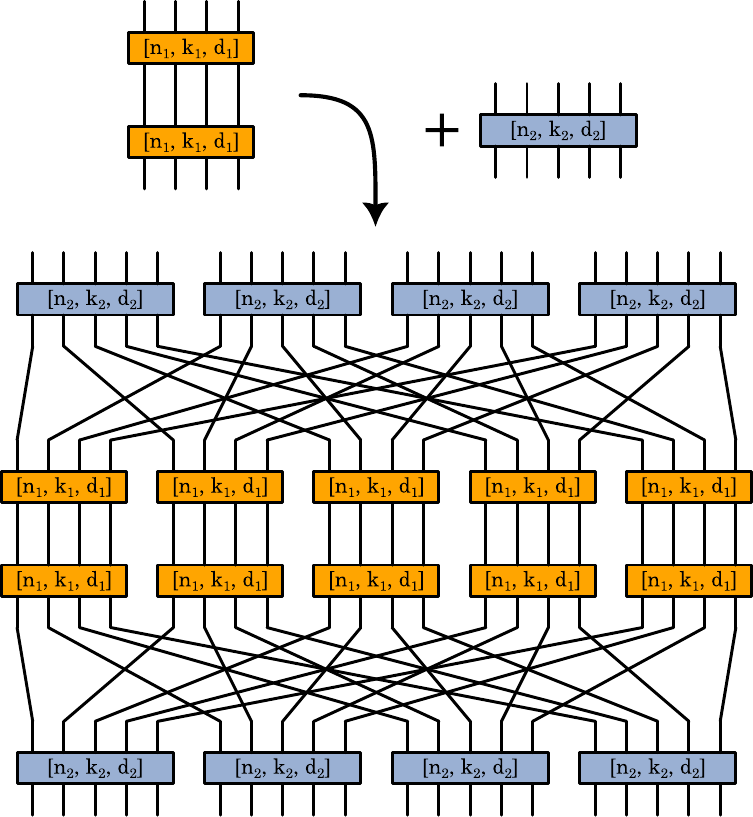}
\caption{Modified concatenation prescription to concatenate a blocklet based on an inner $[n_1, k_1, d_1]$ code with a blocklet based on an outer $[n_2, k_2, d_2]$ code.}
\label{fig:mixedprescription}
\end{figure}

\begin{figure*}[t!]
\centering
\includegraphics[width=0.99\linewidth]{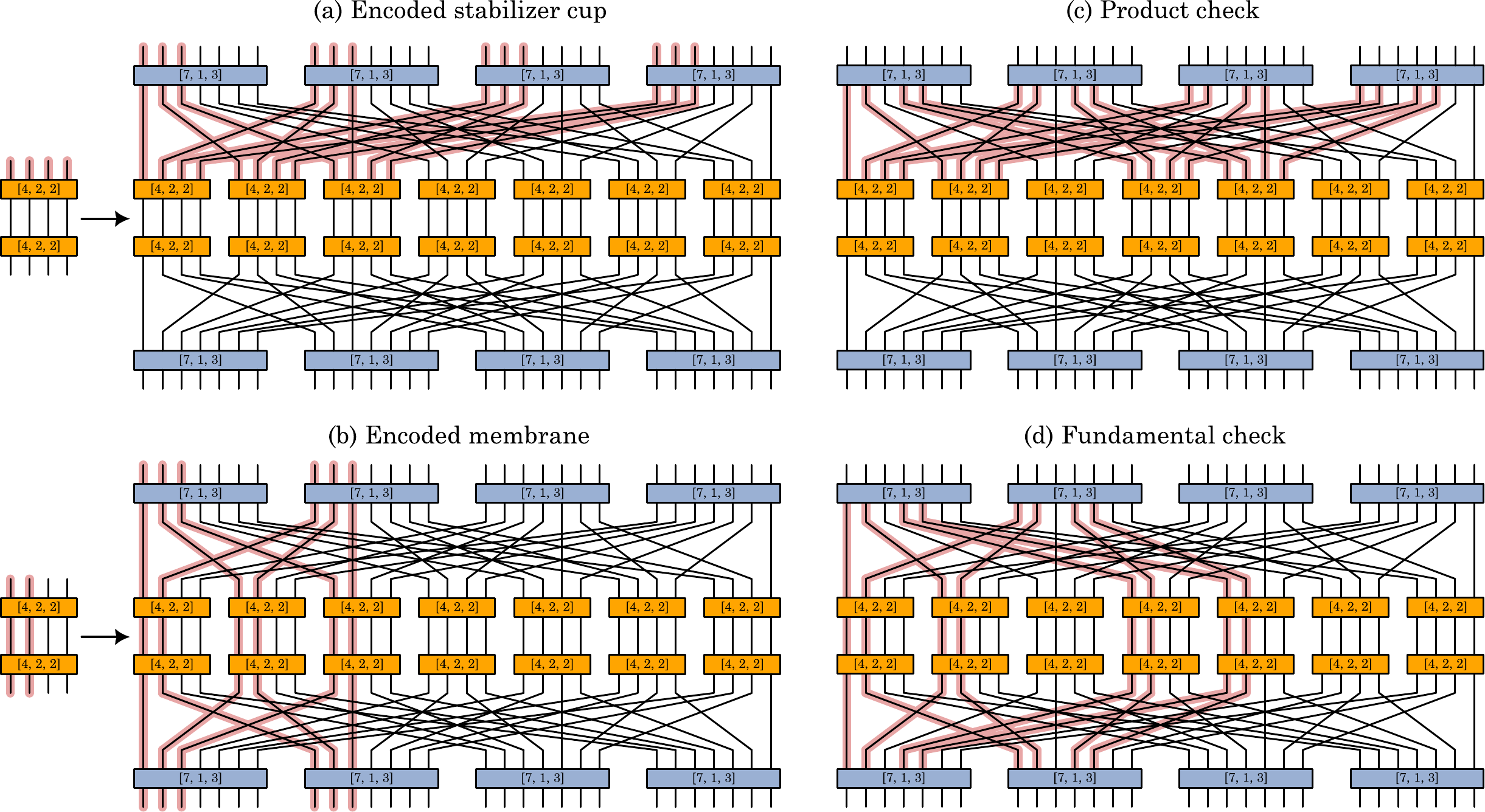}
\caption{Examples of encoded cups, membranes, product checks and fundamental checks in a $[4, 2, 2]$ blocklet concatenated with a $[7, 1, 3]$ blocklets.}
\label{fig:mixedblocklets}
\end{figure*}

\begin{figure}[b!]
\centering
\includegraphics[width=\linewidth]{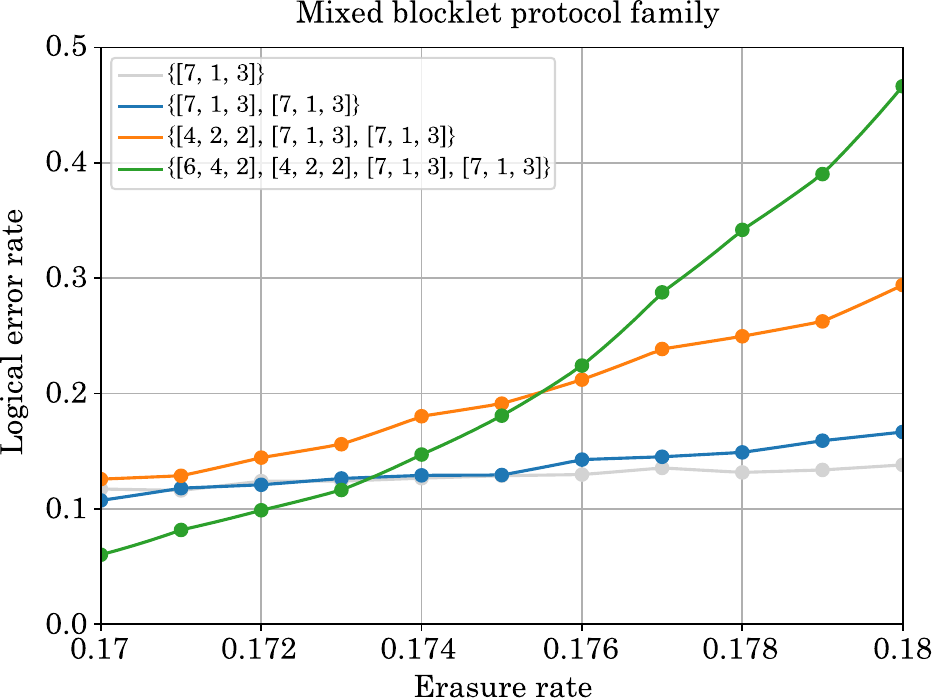}
\caption{Performance of mixed blocklet protocols.}
\label{fig:mixedplot}
\end{figure}

\subsection{Mixed blocklets}

We have described protocols that feature high thresholds and moderate footprint scaling such as the $[5, 1, 3]$ and $[7, 1, 3]$ blocklet protocols, as well as protocols that feature lower threshold but better footprint scaling such as the $[4, 2, 2]$ and $[6, 4, 2]$ blocklet protocols. In fact, one can construct protocols with increasingly better footprint scaling and worse thresholds by using $[n, n-2, 2]$ blocklets. However, one typically would like to have both high thresholds and good footprint scaling in a protocol. To take advantage of both types of protocols, one may therefore want to mix different blocklets in the same protocol.

In Fig.~\ref{fig:mixedprescription}, we describe a modified concatenation prescription that can be used to effectively concatenate a $[n_1, k_1, d_1]$ code with a $[n_2, k_2, d_2]$ code. Rather than replacing individual blocklets, we will now be replacing blocklet pairs with four layers of blocklets. The two central layers each contain $n_2$ blocklets of the $[n_1, k_1, d_1]$ code, whereas the outer layers each contain $n_1$ blocklets of the $[n_2, k_2, d_2]$ code. We will also refer to the $[n_1, k_1, d_1]$ and $[n_2, k_2, d_2]$ codes as the \textit{inner} and \textit{outer} code, respectively. The connections between layers are generated the same way as in the original prescription.

Encoded cups and caps are also obtained as in the original prescription, except now there is one encoded cup and cap for each combination of stabilizers of the inner code and logical operators of the outer code. For example, the $[4, 2, 2]$ code has one $Z$ stabilizer $Z_1Z_2Z_3Z_4$, and the $[7, 1, 3]$ code has one logical $Z$ operator $Z_1Z_2Z_3$. Therefore, when concatenating an inner $[4, 2, 2]$ blocklet with an outer $[7, 1, 3]$ blocklet, an encoded stabilizer cup is formed from cups of $[4, 2, 2]$ blocklets 1, 2, and 3, and membranes of $[7, 1, 3]$ blocklets 1, 2, 3 and 4, as shown in Fig.~\ref{fig:mixedblocklets}a. Similarly, membranes are obtained from the combination of membranes from inner and outer blocklets, such as the one shown in Fig.~\ref{fig:mixedblocklets}b. There are also product checks obtained using the same prescription to combine cups and caps, as the one shown in Fig.~\ref{fig:mixedblocklets}c. One difference is that the fundamental checks that were present in the initial blocklet pair are replaced with fundamental checks that are made from cups and caps of the outer code and membranes of the inner code. Each combination of logical operator of the inner code and stabilizer of the outer code results in one such fundamental check. Since the $[4, 2, 2]$ code has two logical $Z$ operators and the $[7, 1, 3]$ code has three $Z$ stabilizers, this results in six fundamental $Z$ checks (and the same in the $X$ basis), one of which is shown in Fig.~\ref{fig:mixedblocklets}d.

One can even concatenate $[5, 1, 3]$ blocklets with some CSS blocklets, specifically those that are self-dual with respect to their stabilizers. For example, when concatenating $[4, 2, 2]$ inner codes with $[5, 1, 3]$ outer codes, encoded cups, caps, membranes and fundamental checks are obtained the same way as in the CSS prescription. The only difference is in the construction of product checks. Here, we obtain one product check for every pairing of a $[5, 1, 3]$ stabilizer with a $[4, 2, 2]$ stabilizer in one basis. For every $X_{i_1}Z_{i_2}Z_{i_3}X_{i_4}$ stabilizer that is combined with a $Z_{j_1}Z_{j_2}\cdots Z_{j_m}$ stabilizer of a code that is self-dual in its stabilizers (implying that there is also a $X_{j_1}X_{j_2}\cdots X_{j_m}$ stabilizer), blocklets $j_1, j_2, \dots, j_m$ of the $[5, 1, 3]$ code contribute their cups or caps, blocklets $i_1$ and $i_4$ of the CSS code contribute the $X$ versions of their cups or caps, and blocklets $i_2$ and $i_3$ contribute the $Z$ versions of their cups or caps. An example is shown in Fig.~\ref{fig:513mixing}b.

\begin{figure*}[t!]
\centering
\includegraphics[width=0.67\linewidth]{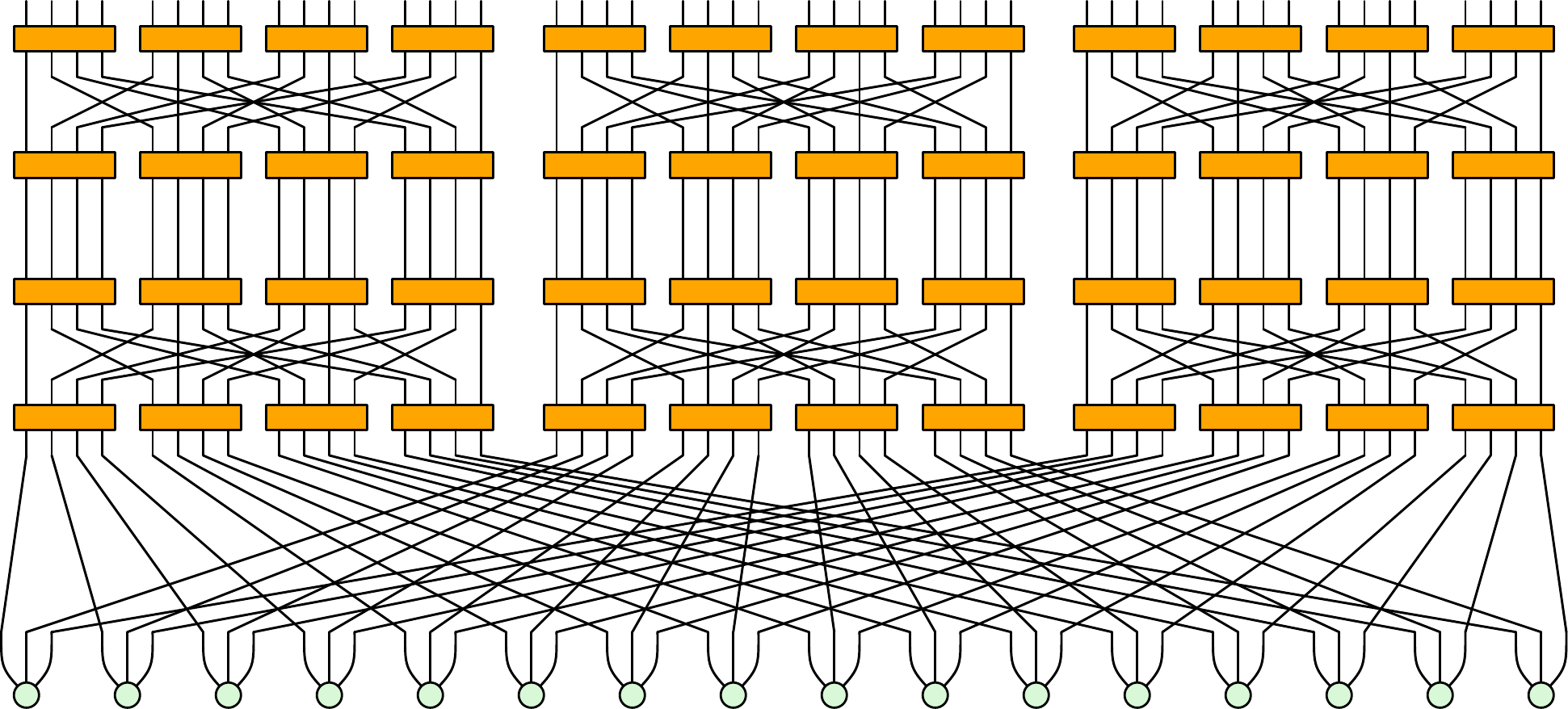}
\caption{Fault-tolerant preparation of a logical 3GHZ with a concatenated blocklet protocol based on an $n=4$ code.}
\label{fig:3ghzblocklet}
\end{figure*}

With the modified concatenation prescription, we can construct a protocol from a list of codes. We first construct a foliated protocol using the first code in the list. We then concatenate each blocklet pair of the protocol with the second code of the list as the outer code. The protocol will now contain blocklet pairs of both the first code and the second code. Each of these blocklet pairs we can concatenate using the third code of the list as the outer code, and so on. One strategy may be to construct protocol families in which we gradually improve the footprint scaling by adding protocols with lower $n/k$ ratios as inner codes. An example of this is shown in Fig.~\ref{fig:mixedplot}. Here, we start by concatenating a $[7, 1, 3]$ code with itself by constructing a $\{[7, 1, 3], [7, 1, 3]\}$ protocol. We then start adding protocols with better scaling by first adding a $[4, 2, 2]$ code as the innermost code, i.e., we construct a $\{[4, 2, 2], [7, 1, 3], [7, 1, 3]\}$ protocol. Finally, we add a $[6, 4, 2]$ code as the innermost code, at which point we are effectively encoding 8 logical qubits with increased error suppression. We find that the two largest protocols have a crossing point at around 17.6\%, which is lower than the threshold of a $[7, 1, 3]$ protocol, but significantly higher than the threshold of the $[6, 4, 2]$ protocol. One can therefore design protocols that balance the desired threshold with the desired footprint scaling.

\section{Logic}
\label{sec:logic}

In the previous section, we focused on protocols for storing logical qubits, i.e., for executing logical identity gates. One can also interpret these identity-gate protocols as fault-tolerant state preparation protocols for logical Bell pairs. To make blocklets useful in the context of fault-tolerant quantum computation, we also need to describe how to perform a universal set of logical operations. As blocklet protocols span a wide range of codes, the exact prescriptions to perform logical operations will depend on the specific codes used in the protocol. Here, we outline general strategies that are applicable to a wide range of codes. The main ingredient of these strategies is the preparation of logical GHZ states in addition to Bell pairs.

\begin{figure}[b!]
\centering
\includegraphics[width=\linewidth]{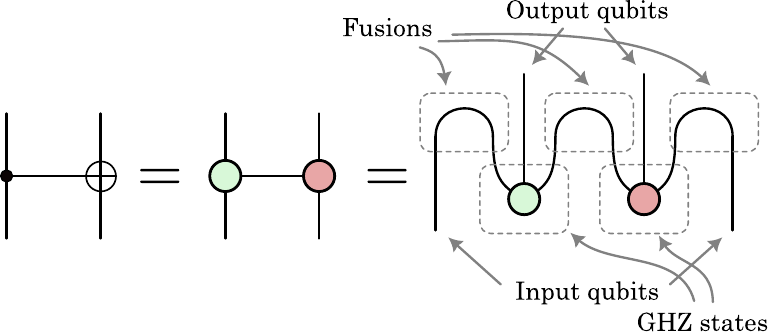}
\caption{A CNOT gate implemented via the preparation of two GHZ states.}
\label{fig:cnotghz}
\end{figure}

\begin{figure*}[t!]
\centering
\includegraphics[width=\linewidth]{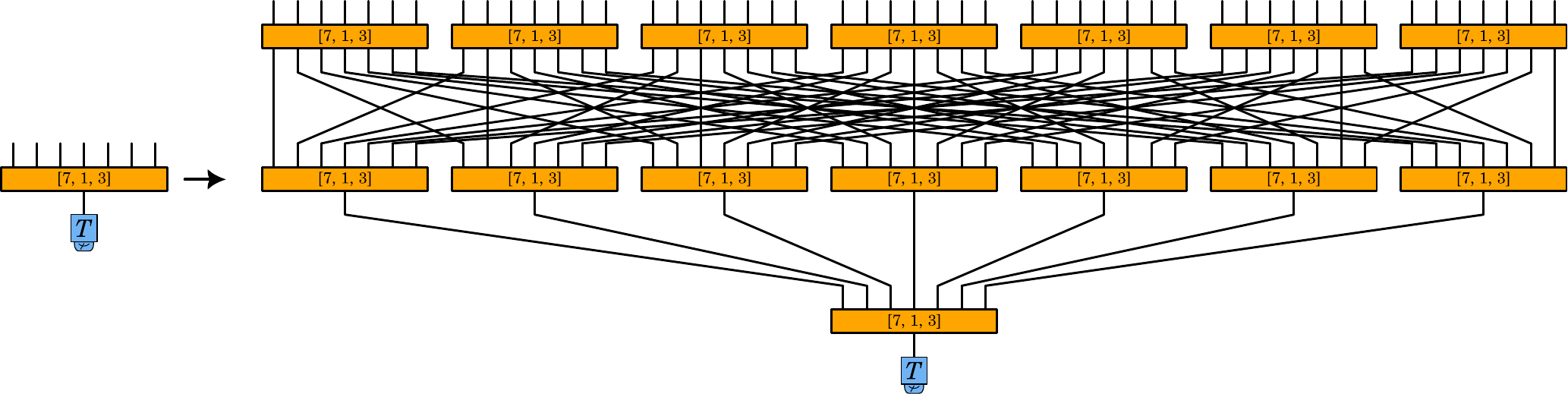}
\caption{State injection in $[7, 1, 3]$ blocklets using a transversal encoding circuit.}
\label{fig:stateinjection}
\end{figure*}

\begin{figure}[b!]
\centering
\includegraphics[width=\linewidth]{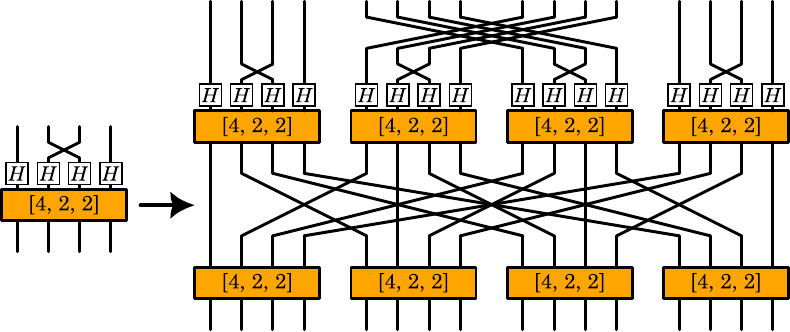}
\caption{Fault-tolerant Hadamard gate of the $[4, 2, 2]$ code and of the $[16, 4, 4]$ code obtained through concatenation.}
\label{fig:422hadamard}
\end{figure}

A commonly used universal gate set is Clifford+$T$. In the language of ZX calculus, any quantum computation can be translated into a ZX diagram that contains phaseless $X$ and $Z$ spiders, Hadamard gates and magic state inputs (and a $|Y\rangle$ state catalyst that may be prepared via distillation at the beginning of the computation). If phaseless spiders are interpreted as GHZ states, then this yields a prescription for the execution of a quantum computation by preparing GHZ states and performing fusions between them. This is not just an arbitrary universal gate set, but indeed the natural set of operations in the active-volume architecture~\cite{Litinski2022}, in which quantum computations are compiled into \textit{logical blocks} and connections between logical blocks. In this context, these logical blocks correspond to logical GHZ states and connections between them to logical fusions. For example, a CNOT gate can be executed by preparing two GHZ states (one of them a Hadamarded GHZ state) and performing fusions between the control qubit, target qubit and the GHZ states, as shown in Fig.~\ref{fig:cnotghz}. With blocklet protocol based on CSS codes, the preparation of logical GHZ states is particularly simple, as it is a transversal operation. As shown in Fig.~\ref{fig:3ghzblocklet} for the example of a blocklet protocol based on an $n=4$ code, this can be achieved by transversally fusing 3GHZ states to three separate code blocks. The operation may also be performed without preparing additional GHZ states by interpreting the GHZ spiders as 3-photon GHZ-state projections, but these have a lower success probability compared to standard type-II fusions, so they may incur some additional overhead.

Single-qubit $X$ and $Z$ measurements are also transversal for CSS codes. The prescription for Hadamard gates will depend on the code. For self-dual codes, it is also transversal, but for other codes one may have a prescription in terms of physical Hadamards and qubit permutations, in which case it can also be extended to concatenated codes. For example, with $[4, 2, 2]$ codes, a Hadamard gate on both encoded qubits may be performed by applying 4 physical Hadamard gates on all qubits and swapping qubits 2 and 3. With a repeatedly concatenated $[4, 2, 2]$ code, this prescription can be applied recursively to each code block, as shown in Fig.~\ref{fig:422hadamard}.

The only remaining operation is state injection for the non-fault-tolerant preparation of $T$-gate magic states that can later be improved through magic state distillation. One way to increase the concatenation level of a qubit is by applying the encoding circuit as a transversal operation. As a state, the encoding circuit is a half-encoded blocklet, i.e., a Bell pair in which one half is encoded in an $[n, k, d]$ code and the other half is a physical qubit. For example, the half-encoded Bell pair of a $[7, 1, 3]$ code is an 8-qubit state as shown in Fig.~\ref{fig:stateinjection}. As a circuit, this corresponds to the encoding circuit that encodes one input qubit into a 7-qubit code. For CSS codes, this operation is a transversal operation, so we can use it to increase or decrease the concatenation level of logical qubits encoded in concatenated codes. In the context of state injection, we want to prepare a physical magic state $T|+\rangle$ and turn it into a logical $T|+\rangle$ state by increasing its encoding to the desired concatenation level. Equivalently, we can perform a $T$ measurement (a $T$ gate followed by an $X$ measurement) on the unencoded qubit, as shown in Fig.~\ref{fig:stateinjection}.

\begin{figure}[b!]
\centering
\includegraphics[width=\linewidth]{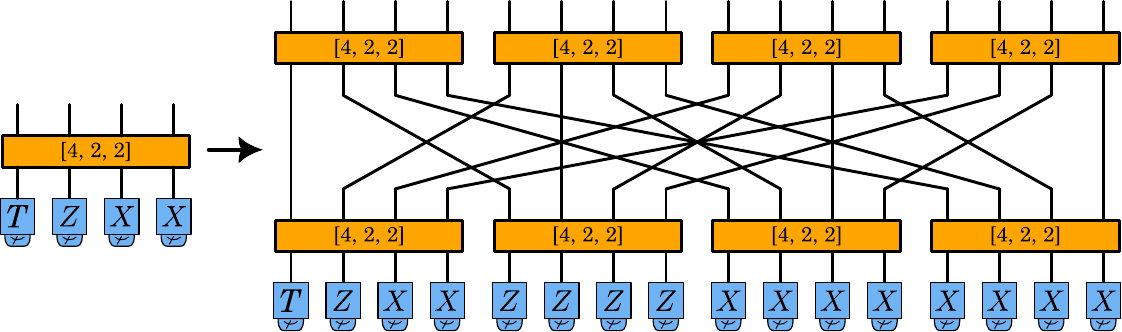}
\caption{State injection in $[4, 2, 2]$ blocklets via single-qubit measurements.}
\label{fig:stateinjection2}
\end{figure}

\begin{figure*}[t!]
\centering
\includegraphics[width=0.99\linewidth]{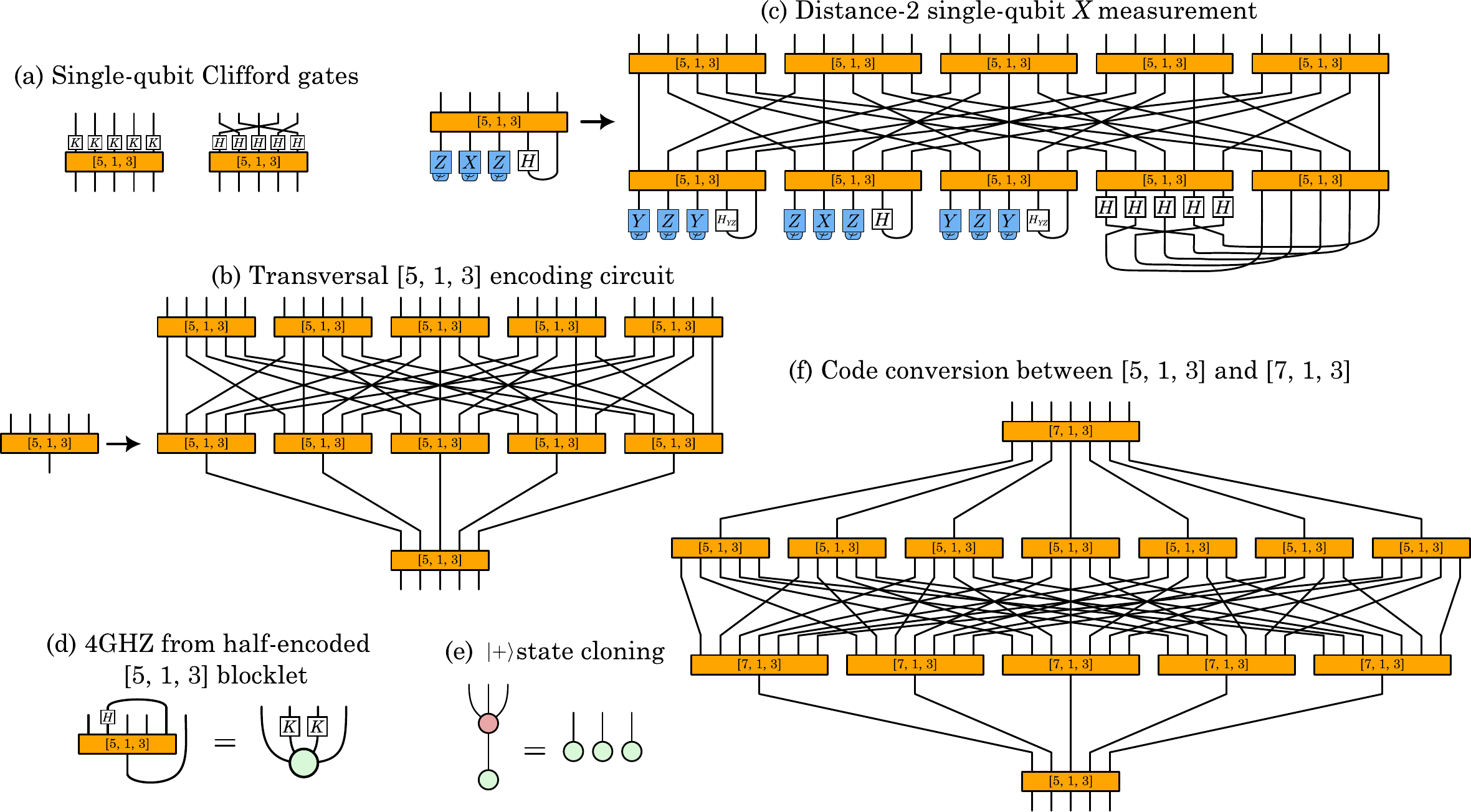}
\caption{Collection of logical operations with $[5, 1, 3]$ blocklets.}
\label{fig:513logic}
\end{figure*}

An alternative to using the encoding circuit is to construct a state injection protocol based on a prescription to unencode a qubit via measurements. For example, one way to turn one of the logical qubits encoded in a $[4, 2, 2]$ code into a physical qubit is by performing a $Z$ measurement on qubit 2, and an $X$ measurement on qubits 3 and 4. The remaining physical qubit 1 will correspond to the unencoded logical qubit. Conversely, we can encode a physical qubit 1 into a $[4, 2, 2]$ code by preparing a $|0\rangle$ state and two $|+\rangle$ states and applying the syndrome readout circuit. If we perform a $T$ measurement on qubit 1 (or if qubit 1 is an unencoded magic state), this is a protocol for state injection. Since it relies on single-qubit Pauli measurements, it can easily be concatenated as shown in Fig.~\ref{fig:stateinjection2}.

GHZ states, Hadamard gates and the preparation of faulty magic states are sufficient for universal quantum computing, as these are all the required ingredients to perform distillation and execute arbitrary logical operations. These are also the native operations of the active-volume architecture~\cite{Litinski2022}.

\textbf{Addressing individual qubits.} One additional operation is required with blocklet protocols that are based on codes with $k > 1$. Suppose that we execute a transversal GHZ-state preparation on three code blocks of a $[4, 2, 2]^L$ protocol. Each code block will store $2^L$ logical qubits, and a transversal GHZ-state preparation will prepare $2^L$ GHZ states at once. If we, e.g., use this to execute CNOT gates, we will always be executing CNOT gates on all $2^L$ logical qubits in the code block. In practice, we want to be able to execute an operation on only one or a specific subset of logical qubits in the code block. We can single out one of the $2^L$ qubits by measuring out all other qubits, turning the GHZ state for all other qubits into a Bell pair (i.e., an identity gate) while retaining the GHZ state for the one qubit that we want to address.

One way of measuring out all but one qubit in a $[4, 2, 2]^L$ code block is by fusing the code block with a code block in which the $2^L - 1$ logical operators that are to be measured are stabilizers. In the case of a $[4, 2, 2]^L$ code, this happens to be a $[4, 1, 2]^L$ code which can be prepared via blocklet concatenation. However, this uses a different resource state and may lead to decreased performance. Alternatively, we can fault-tolerantly measure one of the logical qubits of a $[4, 2, 2]$ code by first adding one more level of concatenation and then measuring out one of the qubits as shown in Fig.~\ref{fig:selectivemeasurement}. This preserves the membranes of one logical qubit (Fig.~\ref{fig:selectivemeasurement}a-b), but measures out the $Z$ membrane of the other logical qubit (Fig.~\ref{fig:selectivemeasurement}c-d) while preserving the code distance, as all fusions and measurements still participate in a check. In a $[4, 2, 2]^L$ protocol, this procedure can be used to measure one of the $2^L$ logical qubits. But we can also concatenate the measurement procedure of Fig.~\ref{fig:selectivemeasurement} by applying the concatenation prescription of Fig.~\ref{fig:prescription}. This yields a measurement protocol that measures two of the four logical qubits of the $[4, 2, 2]^2$ code. Repeated concatenation can be used to obtain measurement prescriptions that measure half of the logical qubits in a $[4, 2, 2]^L$ code. These building blocks enable the efficient measurement of specific subsets of qubits, or the measurement of all but one logical qubit in $L$ measurement steps.

\begin{figure*}[t!]
\centering
\includegraphics[width=0.7\linewidth]{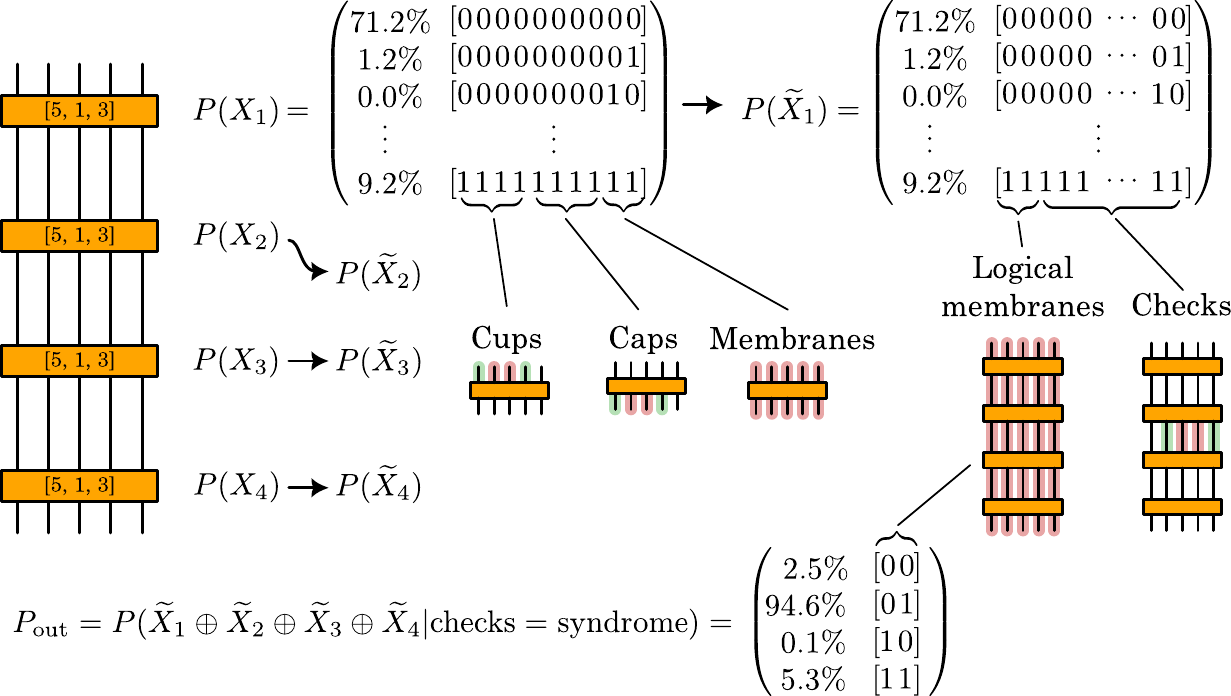}
\caption{Example of a decoding block in a foliated $[5, 1, 3]$ blocklet protocol.}
\label{fig:protocoldecodingblock}
\end{figure*}

\textbf{Logic with $[5, 1, 3]$ blocklets.} A special case are $[5, 1, 3]^L$ protocols, as the $[5, 1, 3]$ code is not a CSS code and therefore does not have transversal GHZ-state projections or even single-qubit Pauli measurements. Single-qubit Clifford gates can be executed either transversally in the case of the $K=SH$ gate, or via additional concatenation-friendly qubit permutations in the case of the Hadamard gate as shown in Fig.~\ref{fig:513logic}a. The $[5, 1, 3]$ code has a known transversal 3-qubit Clifford gate~\cite{Gottesman1998} which can also be interpreted as a transversal projection onto a 6-qubit stabilizer state. Upon closer inspection, this state can be identified as a half-encoded $[5, 1, 3]$ blocklet. In other words, the encoding circuit of a $[5, 1, 3]$ code is a transversal operation of the $[5, 1, 3]$ code and can therefore be used in a concatenated manner as in Fig.~\ref{fig:513logic}b, e.g., for state injection. We can also prepare logical 4-qubit GHZ states (up to single-qubit Clifford gates) by first preparing a logical half-encoded $[5, 1, 3]$ blocklet and then fusing two of the qubits via a logical $XZ/ZX$ fusion (i.e., a fusion preceded by a Hadamard gate), as shown in Fig.~\ref{fig:513logic}d.

While single-qubit Pauli measurements are not transversal, it is possible to construct a concatenation-friendly lower-distance prescription. By measuring qubits 1 and 3 in $Z$, qubit 2 in $X$, and qubits 4 and 5 via an $XZ/ZX$ fusion, we can measure the logical $X$ operator of a $[5, 1, 3]$ code with a code distance of 2 instead of 3. Since this uses single-qubit measurements and fusions, it can be concatenated as shown in Fig.~\ref{fig:513logic}c. This operation can be used to prepare a logical $|+\rangle$ state. However, due to the reduced code distance, higher levels of concatenation (or some post selection) will be required to achieve a sufficiently low error rate, so the preparation of $|+\rangle$ states in this manner may be very expensive. Fortunately, it is only necessary to prepare a high-quality $|+\rangle$ state once at the beginning of the quantum computation. Additional $|+\rangle$ states can be generated by fusing them to Hadamarded GHZ states as shown in Fig.~\ref{fig:513logic}e, spending one 4GHZ state to turn one $|+\rangle$ into three $|+\rangle$ states. One may then keep a high-quality $|+\rangle$ state as a catalyst state to generate additional high-quality $|+\rangle$ states on demand. These can be used both for logical single-qubit measurements and to turn 4-qubit GHZ states into 3-qubit GHZ states.

These ingredients are sufficient for universal quantum computation with $[5, 1, 3]$ blocklets. One may also explore additional constructions for logical operations. For example, $[5, 1, 3]$ codes can be fault-tolerantly converted into $[7, 1, 3]$ codes via the protocol shown in Fig.~\ref{fig:513logic}f, which may be used to take advantage of the transversal gates of the $[7, 1, 3]$ code. It is also possible to prepare 6GHZ states instead of 4GHZ states by preparing logical fully-encoded blocklet states instead of half-encoded ones. There also exist constructions for non-transversal fault-tolerant gates for the $[5, 1, 3]$ code~\cite{Yoder2016}.

\section{Decoding}
\label{sec:decoding}

The simulation results presented so far relied on an erasure-only error model, as erasures can be optimally decoded with an efficient general-purpose decoder. This is no longer the case in the presence of Pauli errors. Therefore, an efficient well-performing decoding scheme is required to make blocklet protocols useful in practice. In this section, we describe a decoding scheme that we refer to as \textit{hierarchical decoding}. It is not optimal and several improvements over what is described in this section are possible, so it should be interpreted as a proof of principle demonstrating that efficient decoding with acceptable performance is possible, rather than a statement about the achievable performance.

The decoder works by keeping track of \textit{Pauli web distributions} (PWDs) of blocklets. Pauli errors affecting the qubits of an $n$-qubit blocklet will flip some subset of its $n$ Pauli webs. Our goal is to keep track of the likelihood of different flip patterns. Therefore, a PWD is a probability distribution $P(X)$ of a binary $n$-bit random variable $X$, which can be represented as a list of $2^n$ probabilities. For example, for a $[5, 1, 3]$ blocklet with 10 Pauli webs, it is a list of $2^{10} = 1024$ probabilities, each describing the likelihood of one of the $1024$ possible configurations of flipped Pauli webs.

We partition our decoding problem into \textit{decoding blocks}. Each decoding block has a number of input PWDs $P(X_i)$ and generates one output PWD $P(X_{\rm out})$. For example, consider a foliated $[5, 1, 3]$ blocklet protocol with 4 foliation layers at $L=1$. In this case, there is only one decoding block, which is shown in Fig.~\ref{fig:protocoldecodingblock}. There are four input PWDs, one for each blocklet. Each input PWD has $n_{\rm in} = 10$ Pauli webs. The output PWD is one that keeps track of only $n_{\rm out} = 2$ Pauli webs which are the logical $X$ and $Z$ membranes of the protocol, i.e., $X_{\rm out}$ is a 2-bit variable in this decoding block. Furthermore, the decoding block contains $n_{\rm checks} = 16$ checks, four from each pair of neighboring blocklets (with periodic boundary conditions).

\begin{figure}[t!]
\centering
\includegraphics[width=\linewidth]{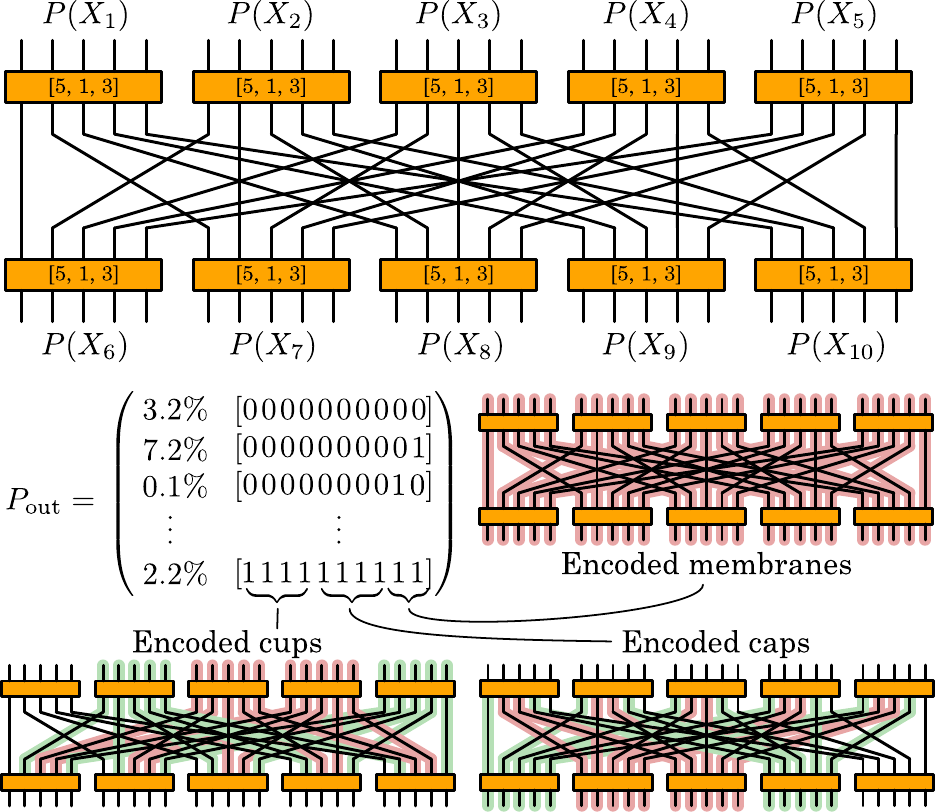}
\caption{Example of a concatenated decoding block in a $[5, 1, 3]$ blocklet protocol.}
\label{fig:concatenateddecodingblock}
\end{figure}

To assemble $P(X_{\rm out})$ from the input PWDs $P(X_i)$, we first convert every $P(X_i)$ into $P(\widetilde{X}_i)$ where $\widetilde{X}_i$ are binary variables with $n_{\rm out} + n_{\rm checks}$ bits. To convert $X_i$ to $\widetilde{X}_i$, we assemble a matrix $C_i$ with $n_{\rm out} + n_{\rm checks}$ rows and $n_{\rm in}$ columns. Each row corresponds to one check or output web, and each column corresponds to one input web of the $i$-th input blocklet. Each check or output web is assembled from blocklet Pauli webs (see Fig.~\ref{fig:blockletintro}), so every input Pauli web may contribute to some subset of checks and output webs. An entry in $C_i$ in a row corresponding to a check or output web $a$ and column corresponding to an input web $b$ is only 1, if $b$ contributes to $a$, else it is 0. We then convert $X_i$ to $\widetilde{X}_i$ by computing $\widetilde{X}_i = C_i X_i$. In our example protocol, each column in $C_i$ contains only a single 1, as each cup and cap contributes to only one check and each blocklet membrane contributes to one logical membrane. The converted PWDs $P(\widetilde{X}_i)$ are typically sparse. Rather than describing which input blocklet Pauli webs are flipped, they describe which checks and output Pauli webs are flipped as a result.

\begin{figure}[t!]
\centering
\includegraphics[width=\linewidth]{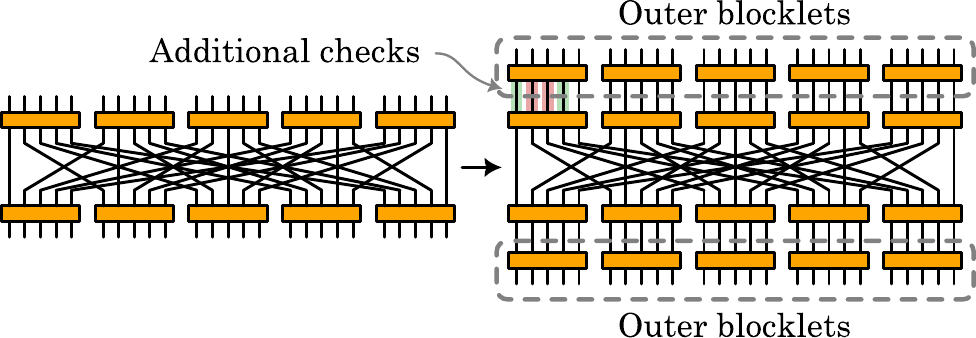}
\caption{Example of an extended decoding block in a $[5, 1, 3]$ blocklet protocol.}
\label{fig:extendeddecodingblocks}
\end{figure}

\begin{figure*}[t!]
\centering
\includegraphics[width=\linewidth]{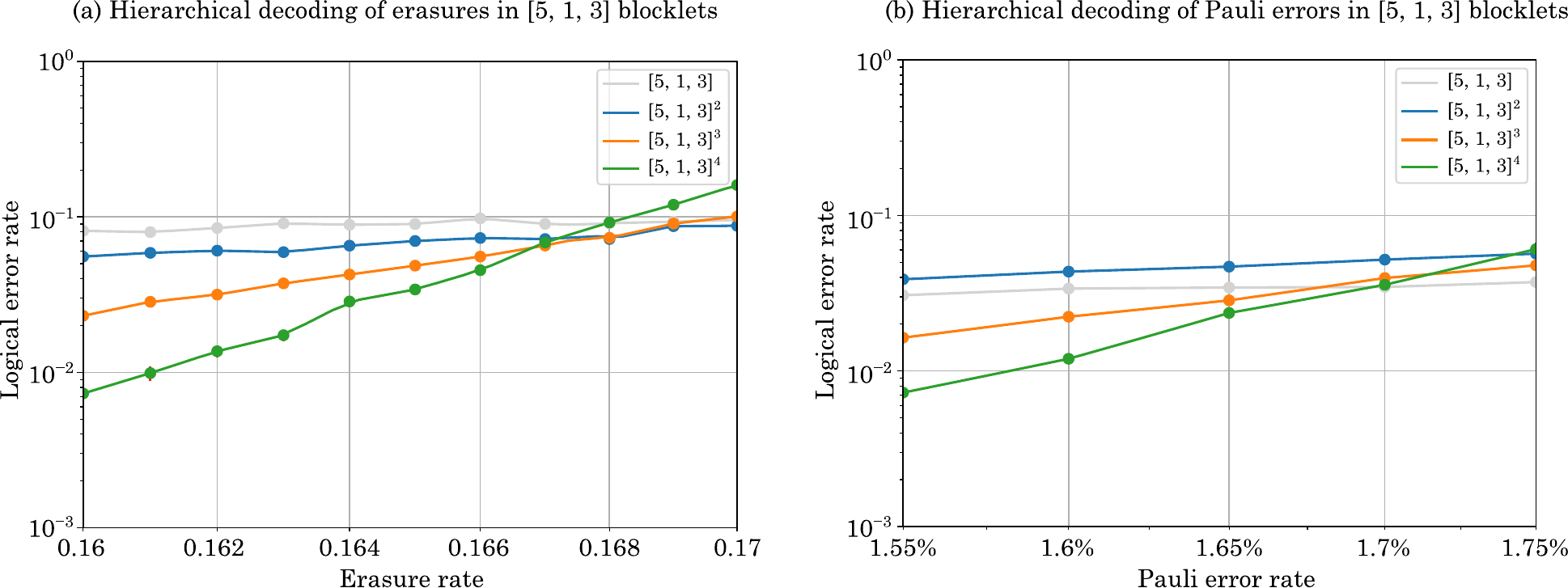}
\caption{Monte Carlo simulation results of a $[5, 1, 3]$ blocklet protocol with a hierarchical decoder.}
\label{fig:hierarchicalplots}
\end{figure*}

Next, we compute $P(\bigoplus_i \widetilde{X}_i)$ where $\oplus$ is a bitwise XOR operation, which yields a probability distribution that tells us for each output web and check whether it was flipped an even or odd number of times due to the flipped input webs of the input blocklets. $P(X_1 \oplus X_2)$ can be computed from two probability distributions $P(X_1)$ and $P(X_2)$ of $n$-bit binary random variables $X_1$ and $X_2$ represented as sparse vectors of $m = 2^n$ numbers with sparsities $s_1$ and $s_2$ either with $\mathcal{O}(m \log m)$ operations via a fast Walsh-Hadamard transform or with $\mathcal{O}(s_1 s_2)$ operations by pairing up all nonzero probabilities. We condition the resulting probability distribution on the checks matching the measured syndrome, i.e., we eliminate all entries from the probability vector in which the pattern of flipped checks does not match the observed syndrome, before normalizing the probability distribution. The result is $P(X_{\rm out})$. In the case of the foliated $[5, 1, 3]$ protocol in Fig.~\ref{fig:protocoldecodingblock}, the output PWD tells us which logical membranes are likely to have been flipped by errors. We can then correct the error by flipping fusion outcomes to selectively flip logical membranes. A logical membrane of a logical operator $X_i$ (or $Z_i$) can be flipped by flipping the $XX$ (or $ZZ$) fusion outcomes of all fusions in one layer of the protocol (say, the first layer) that are part of $Z_i$ (or $X_i$).

\textbf{Decoding blocks of concatenated protocols.} Our strategy for decoding an $L=2$ protocol will be to map it back to the decoding problem of an $L=1$ protocol. We will assign a \textit{level} to each decoding block. For an $L=1$ protocol, there is only one decoding block and it is assigned a level 1. To construct the decoding blocks of the $L=2$ protocol, we will keep the existing decoding block, but increase its level to 2.
The inputs to the level-2 decoding blocks will be generated from outputs of level-1 decoding blocks. We introduce decoding blocks at level 1 that have the same structure as concatenation blocks, i.e., for $[n, k, d]$ blocklets, they consist of $2n$ input blocklets and generate one output blocklet whose output webs are the encoded webs of a decoding block. An example for a concatenated $[5, 1, 3]$ blocklet protocol is shown in Fig.~\ref{fig:concatenateddecodingblock}. Here, the checks are the product checks inside the concatenated block. The decoding block is executed the same way as described previously: We convert $P(X_i)$ to $P(\widetilde{X}_i)$ with the help of conversion matrices $C_i$, compute an XOR distribution $P(\bigoplus_i \widetilde{X}_i)$, condition it on the checks matching the syndrome, and obtain an output PWD $P(X_{\rm out})$. There will be four level-1 decoding blocks in our examples, the outputs of which are used as inputs in the level-2 decoding block. For higher levels of concatenation, we level up all decoding blocks, before again introducing concatenated level-1 decoding blocks as in Fig.~\ref{fig:concatenateddecodingblock}. Each level-$i$ decoding block will produce an output PWD that is used as an input in a level-$(i+1)$ decoding block. Therefore, a protocol with concatenation level $L$ is decoded in $L$ steps, where in each step we execute all decoding blocks of one level, before proceeding with the decoding blocks of the next level. The lowest level will contain the most decoding blocks, but all decoding blocks of one level can be executed in parallel.

\begin{figure*}[t!]
\centering
\includegraphics[width=0.98\linewidth]{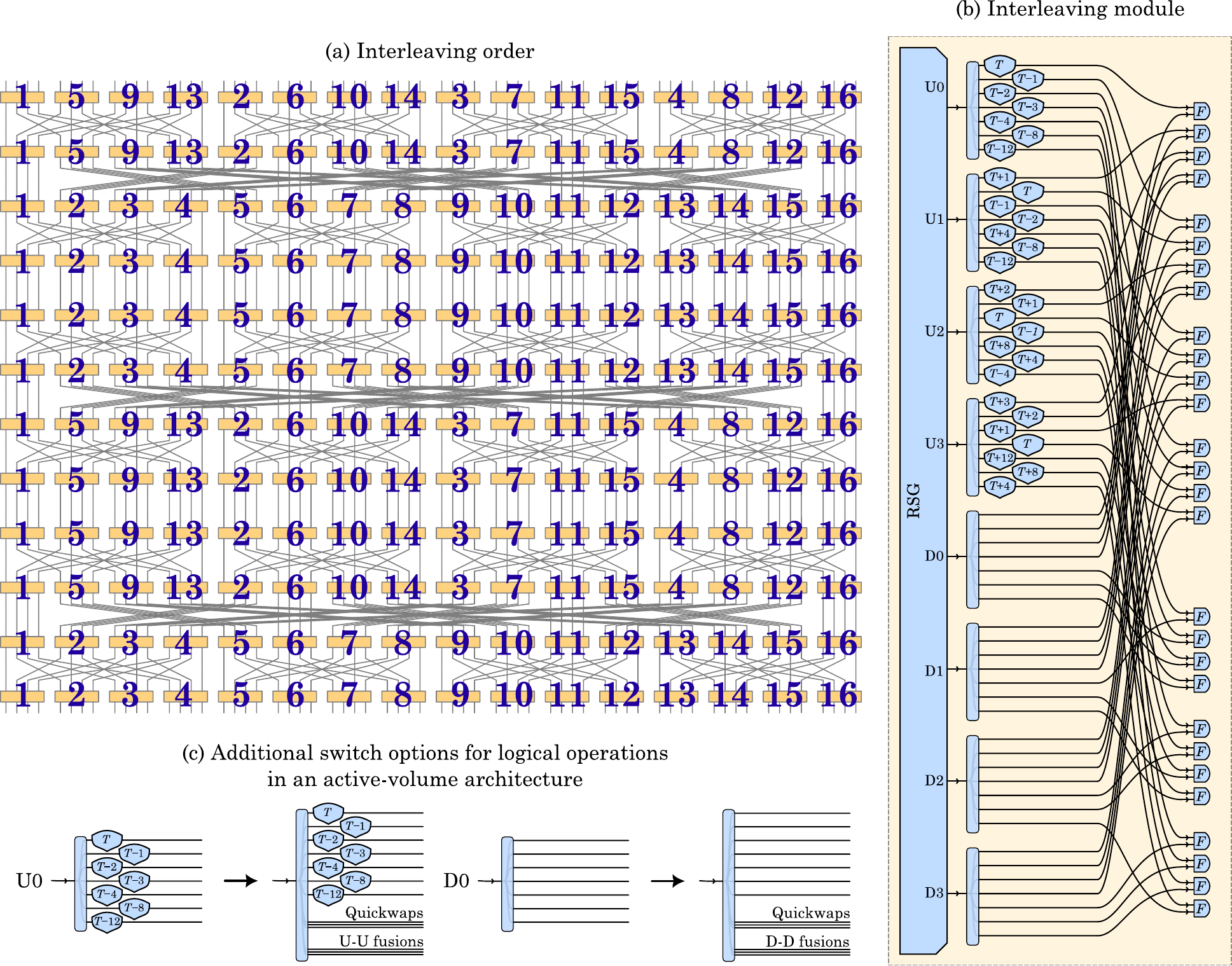}
\caption{Interleaving order and interleaving modules for an $L=3$ protocol based on a 4-qubit blocklet resource state with $T=16$ resource states per layer.}
\label{fig:interleaving}
\end{figure*}

\textbf{Extended decoding blocks.} We will perform one additional modification to our decoding blocks. We will also incorporate some of its environment into each decoding block, i.e., we consider additional input blocklets in each decoding block. In the simplest case, this is just the neighboring blocklet of each of the original input blocklets. We will refer to these additional blocklets as \textit{outer blocklets} as shown in Fig.~\ref{fig:extendeddecodingblocks}. Outer blocklets do not contribute to the output webs of the decoding block. However, they do contribute to the additional checks that are introduced when extending the decoding blocks, e.g., the additional fundamental checks in Fig.~\ref{fig:extendeddecodingblocks}. When computing the output blocklet PWD, we will also condition on these checks.

\textbf{Simulation results.} In Fig.~\ref{fig:hierarchicalplots}, we present Monte Carlo simulation results for a concatenated $[5, 1, 3]$ blocklet protocol family that is decoded using a hierarchical decoder, i.e., the same protocol as in Fig.~\ref{fig:erasureplots}b but with a different decoder. For an erasure-only error model, we find that the lines of the $L=3$ and $L=4$ protocol cross at an erasure rate of 16.7\% instead of 19.1\%. In other words, this decoder is not optimal and in this case appears to underperform by 13\% compared to the optimal decoder. However, this decoder can also be used in the presence of Pauli errors. For a Pauli-only error model, we find that the lines of the $L=3$ and $L=4$ protocols cross at a 1.7\% Pauli error rate.

\textbf{Possible improvements.} The simulation results demonstrate that a hierarchical decoder can decode both Pauli errors and erasures, but additional performance improvements are possible. One may explore adding more outer blocklets to the decoding blocks, or using larger decoding blocks altogether, e.g., by partitioning the decoding problem into blocklet pairs rather than individual blocklets. Another potential improvement could be to compute more accurate input PWDs prior to executing the decoding blocks in each level via more general methods for the computation of marginal probability distributions such as belief propagation. It may also be possible to devise an iterative decoding scheme in which PWDs are refined in subsequent rounds of decoding by using the output PWDs from previous decoding rounds. Finally, to make the hierarchical decoding scheme more applicable to blocklet protocols with $k > 1$, one may also need to develop methods to further subpartition the decoding problem. With blocklet protocols based on $[n, k, d]$ codes with $k>1$, the number of output Pauli webs of each decoding block grows by a factor of $k$ in each level. The sizes of PWDs may therefore become very large for large $k$ or after a large number of concatenation steps, necessitating the use of probability cutoffs or other truncation methods that can reduce the accuracy of the decoder, so avoiding very large PWDs is crucial. Generally, when optimizing the performance of a decoder, one should be mindful that it is not primarily the threshold that matters, but rather the performance at a specific sub-threshold operating point.

\section{Photonic implementation}
\label{sec:photonic}

In this section, we outline how blocklet protocols can be implemented with photonic hardware, specifically in an interleaving architecture as introduced in Ref.~\cite{Bombin2021}. In an interleaving architecture, a quantum computer is partitioned into modules, where each module is a collection of silicon-photonic chips that functionally implement a resource-state generator, switches, delay lines and fusion-measurement devices. A resource-state generator (RSG) outputs a resource state in every time step (RSG cycle). In the case of blocklet protocols, the resource state is a $2n$-qubit blocklet, where we label the qubits corresponding to the top ports as $U_0\dots U_{n-1}$ and the qubits corresponding to the bottom ports as $D_0\dots D_{n-1}$. Each qubit enters a switch that routes it through one of several delay lines which delay the qubit for a fixed number of RSG cycles. Pairs of qubits ultimately arrive at a fusion device that performs a fusion measurement.

Each interleaving module can be thought of as one logical qubit, or one code block in the case of $k>1$. While our blocklet protocol construction specifies how many resource states need to be generated and how they need to be fused, it does not specify the order in which we should generate them. As each interleaving module contains only one RSG, it will generate all resource states sequentially. We need to label our resource states to determine the order in which we will generate them. A good labeling is one in which there is only a small number of label differences between fused resource state. If a resource state is generated at time $t_1$ and another resource state that it is fused with is generated at time $t_2$, then an interleaving module needs to support a delay line that can effectively delay qubits for a time $t_2 - t_1$. We are interested in minimizing the number of different delay lengths, as this minimizes the number of switch options and therefore the size of switches.

In topological codes such as surface codes, we can construct interleaving modules that use a constant number of delay-line lengths regardless of the code distance. With concatenated blocklet protocols, the price that we pay for improved performance is that the non-locality of the connections grows with each concatenation step. While we cannot implement interleaving modules with a constant number of delay lengths, we can find constructions that use $\mathcal{O}(\log d)$ (or $\mathcal{O}(L)$) many different delay lengths.

We will generate resource states layer by layer. In a $[n,k,d]^L$ protocol, each layer will contain $n^{L-1}$ resource states. We start by labeling resource states in the first layer in ascending order. In every subsequent layer, we will either copy the labels from the previous layer if it is an odd layer, or assign each blocklet the label of its partner blocklet (which is the blocklet that is transversally fused to it) if it is an even layer. Remember that resource states correspond to blocklet pairs that are composed of blocklets and their partner blocklets as in Fig.~\ref{fig:collapse}, so they must be assigned the same label. An example of that is shown in Fig.~\ref{fig:interleaving}a for an $L=3$ protocol with blocklet resource states with $n=4$. 

We find that with this assignment of labels, each qubit $U_j$ may need to be delayed for $T + n^i \cdot (j + \{0, 1, 2, \dots, n-1\})$ RSG cycles, where $T=n^{L-1}$ and $i \in \{0, 1, \dots, L-2 \}$. In other words, the total number of switchable delay lengths is $(L-1) \cdot (n-1) + 1$. For the example with $n=4$ and $L=3$, this implies 7 different switchable delay lengths per qubit, as shown in Fig.~\ref{fig:interleaving}b. While the number of switch options is no longer constant, the number of switch options in practice is still compatible with photonic hardware.

The switch options discussed thus far are sufficient for logical identity operations. Additional switch options are required for other logical operations. The specific requirements will depend on the methods that are used for logical operations, but in general we may use a construction that is almost identical to the active-volume interleaving modules in Ref.~\cite{Litinski2022}. This requires the addition of $\mathcal{O}(\log N)$ switch options for quickswaps and transversal fusions ($U$-$U$ and $D$-$D$ fusions) between different code blocks as shown in Fig.~\ref{fig:interleaving}c, where $N$ is the total number of logical qubits in the quantum computer. Depending on the chosen methods for logical operations, one may also need at least one additional switch option for the preparation of logical GHZ states.

\section{Conclusion}

In this paper, we have introduced a very versatile toolbox for the construction of fault-tolerant protocols based on concatenation and transversal gates. Some of the protocols that we constructed have relatively high erasure thresholds, such as the $[7, 1, 3]$ and $[5, 1, 3]$ blocklet protocols, whereas others have lower thresholds but better footprint scaling as a function of the code distance, such as the $[4, 2, 2]$ and $[6, 4, 2]$ blocklet protocols. By mixing blocklets with different protocols, e.g., by initially concatenating with high-threshold codes before gradually adding lower-threshold code with better footprint scaling, one may benefit from the properties of both types of protocols. We find that in the context of FBQC, these protocols perform remarkably well in comparison to surface codes. It is encouraging to see that good fault-tolerant protocols can be found in parameter regimes that are very different from those of surface codes, as one of the most promising protocols, the $[5, 1 ,3]$ blocklet protocol family, is non-topological, non-LDPC and non-CSS.

While we have shown results for some protocols, many interesting protocols likely remain unexplored. This includes protocols based on other codes and different mixtures of codes. One may also explore different ways of partitioning the protocols into resource states. While we have focused on the performance of these protocols in the context of FBQC, they can all straightforwardly be interpreted as quantum circuits, which means they may also be interesting in the context of circuit-based quantum computing. However, their performance may differ with a circuit-based error model, so additional simulations are necessary to determine if blocklet protocols are also useful in CBQC, assuming that the hardware can support the necessary connectivity. As the circuits corresponding to blocklet protocols contain only low-weight measurements, they appear promising in a CBQC setting.

One may also explore variations of the protocol construction described in this paper. It is relatively straightforward to extend this construction to blocklets based on two different codes, i.e., encoded Bell pairs with two different codes on both sides. This can be used to construct an even wider range of protocols from essentially any (CSS) stabilizer state. After all, any phaseless ZX diagram has Pauli webs that can be interpreted as cups, caps and membranes, implying that any phaseless ZX diagram can be used as a blocklet. It may also be useful to explore protocol constructions that allow for more fine-grained control over the code distance. One drawback of concatenation is that each concatenation step leads to big jumps in the code distance, whereas topological constructions can be used to increase the code distance in small steps. However, in the context of photonic FBQC, this is not necessarily a problem in practice, as more granular control control can be achieved by other means, such as by adjusting the local encoding or the post-selection overhead in resource-state generation. Moreover, proving (or disproving) the distance scaling conjecture also remains an open problem.

We have presented different approaches for logical operations with blocklets, but many open questions remain. For one, these operations rely on the preparation of magic states, so it is desirable to construct magic-state preparation protocols that minimize their cost. Furthermore, logical operations on blocklet protocols with $k>1$ act on multiple groups of qubits at the same time. We described operations for blocklets with $k>1$ that can be used to selectively address qubits, but perhaps one can also take advantage of the ability to execute multiple operations in parallel on certain subgroups of qubits in a blocklet quantum computation.

Finally, it would be interesting to develop improved decoding schemes for blocklet protocols, e.g., via some of the suggested improvements to the hierarchical decoder that are described at the end of Sec.~\ref{sec:decoding}. Improvements may not only concern the performance in terms of logical error rate, but also in terms of mapping to hardware and runtime in practice. The hierarchical decoder is highly parallelizable, so may have advantageous properties when implemented using dedicated hardware.

\section*{Acknowledgments}

I would like to thank Kwok Ho Wan for pointing out important inconsistencies between the claimed protocol distances in an earlier draft and the observed subthreshold behavior. I would also like to thank Jake Bulmer, Andrew Doherty, Terry Farrelly, Naomi Nickerson, Sam Roberts, Terry Rudolph, Kwok Ho Wan and Dominic Williamson for helpful feedback on the draft, and my colleagues in the PsiQuantum architecture team for many helpful discussions and their support with simulation software tools.

\appendix

\section{Conjecture about the code distance of concatenated blocklet protocols}
\label{app:distance}

While the code distance of a repeatedly concatenated $[n, k, d]$ code is $d^L$ for $L$ levels of concatenation, determining the code distance of an $[n, k, d]^L$ protocol (i.e., the lowest-weight error string that is undetectable by the checks of the protocol) is not straightforward. In this section, we present a general construction for error strings that have a weight of $d_{\rm prod} \cdot d^{L-2} \leq d^L$. We conjecture these to be the lowest-weight error strings of the protocol. We find support for this conjecture in the below-threshold scaling of the logical error rate of different blocklet protocols, although a proof for this conjecture remains an open problem.

\begin{figure}[b!]
\centering
\includegraphics[width=0.6\linewidth]{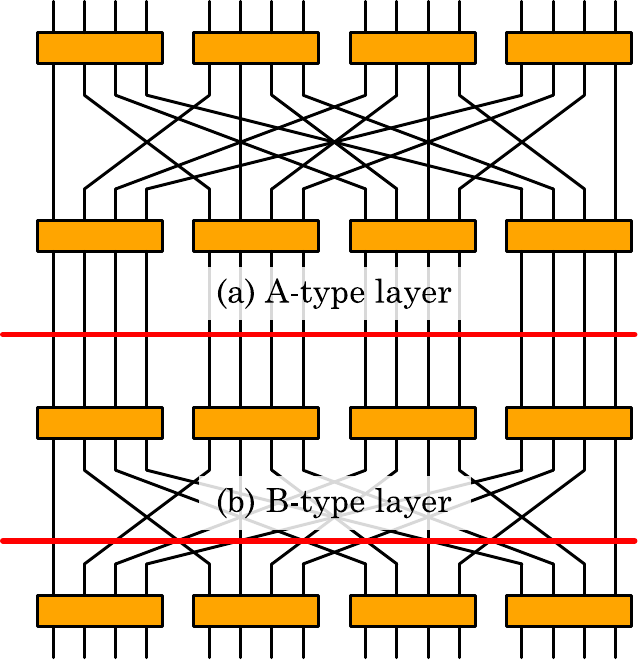}
\caption{Two types of layers in an $L=2$ protocol.}
\label{fig:distancelayers}
\end{figure}

\begin{figure*}[t!]
\centering
\includegraphics[width=\linewidth]{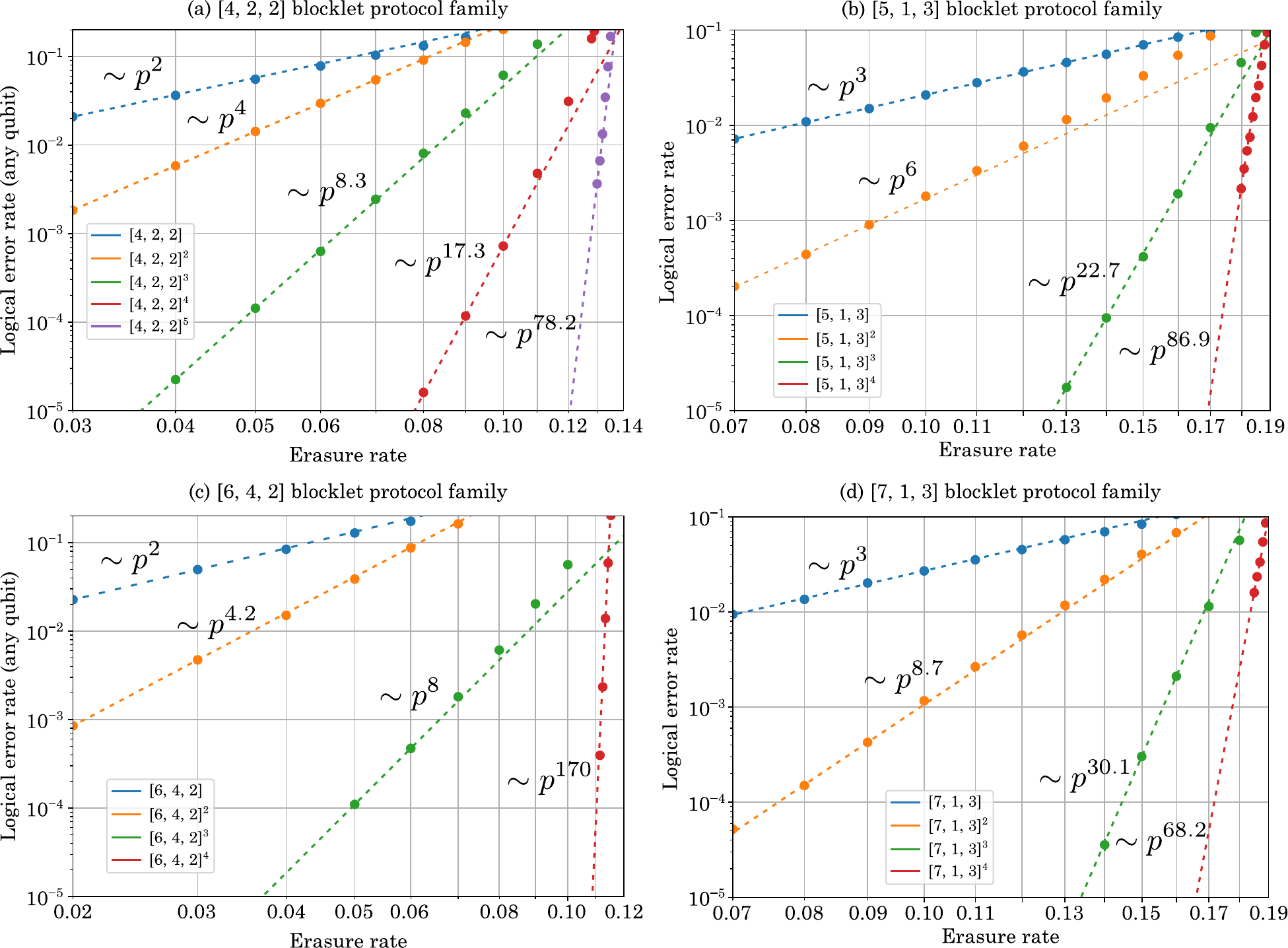}
\caption{Subthreshold behavior of blocklet protocols based on lowest-logical-error-rate points from Monte Carlo sampling.}
\label{fig:subthreshold}
\end{figure*}

Consider an $L=2$ protocol. It contains two types of layers that we will refer to as \textit{A-type} and \textit{B-type} layers as shown in Fig.~\ref{fig:distancelayers}. We are interested in finding the lowest-weight error configurations within a layer that flip a logical membrane without flipping any checks. The fusion outcomes in each layer are encoded in an error-correcting code. In other words, if errors were restricted to the fusions in a single layer, the corresponding fusion-outcome bits would be encoded in a code whose parity checks correspond to the checks of the blocklet protocol constrained to this layer. Suppose that the $Z$ checks of the $[n, k, d]$ code are described by the parity check matrix $H_Z$ and its $Z$ logical operators by $L_Z$. Then the $ZZ$ fusion outcomes in A-type layers are encoded in the $[n, k, d]$ code concatenated with itself, i.e., the parity checks are described by $\mathbbm{1}_n \otimes H_Z$ and $H_Z \otimes L_Z$, and the logical operators by $L_Z \otimes L_Z$. (The same applies to $X$ checks, $X$ logicals and $XX$ fusion outcomes.) If $E_Z$ is a logical error string of $H_Z$, then its weight is at least $d$. The logical error strings of the concatenated code are $E_Z \otimes E_Z$, and therefore have a weight of at least $d^2$.

However, in B-type layers, the fusion outcome bits are encoded in a different code. We will refer to this code as the \textit{product code} of the $[n, k, d]$ code. Its parity checks are described by $H_Z \otimes H_Z$, $H_Z \otimes L_Z$ and $L_Z \otimes H_Z$, which correspond to the product checks, encoded cups and encoded caps of the blocklet protocol, respectively. The logical operators are still $L_Z \otimes L_Z$. While $E_Z \otimes E_Z$ are weight-$d^2$ logical error strings of this code, a product code may also have lower-weight errors $E_{\rm prod}$. Therefore, the distance of an $L=2$ protocol will be limited by the distance $d_{\rm prod}$ of this product code.

\begin{figure*}[t!]
\centering
\includegraphics[width=\linewidth]{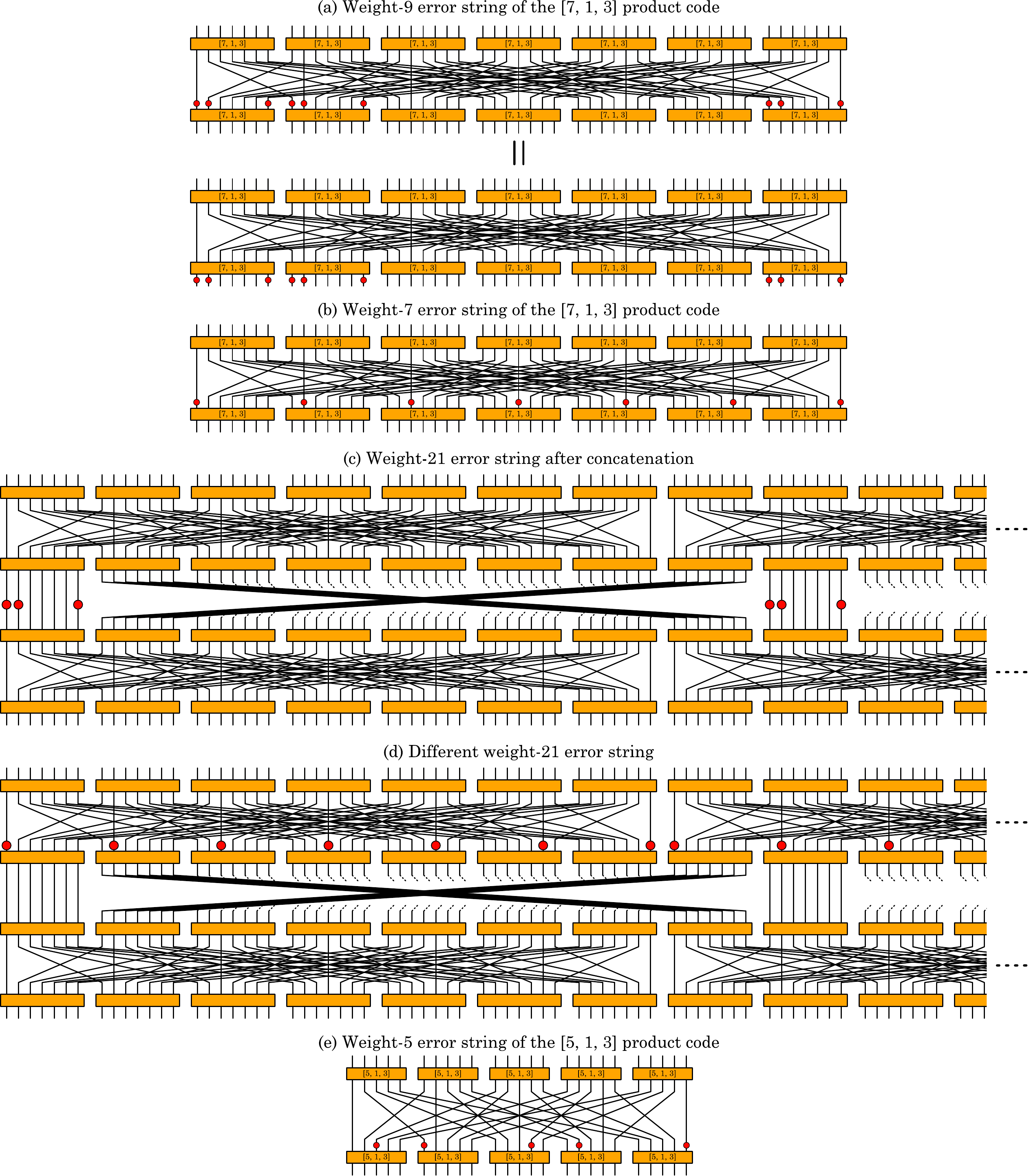}
\caption{Lowest-weight error strings of different product codes.}
\label{fig:productcode}
\end{figure*}

Consider the example of a $[7, 1, 3]$ code whose $Z$ (or $X$) parity check matrix is
\begin{equation}
H_{[7,1,3]} = \begin{pmatrix}
1 & 1 & 1 & 1 & 0 & 0 & 0 \\
0 & 0 & 1 & 1 & 1 & 1 & 0 \\
0 & 1 & 0 & 1 & 0 & 1 & 1 \\
\end{pmatrix} \, ,
\end{equation}
and whose $Z$ (or $X$) logical operator is
\begin{equation}
L_{[7,1,3]} = \begin{pmatrix}
1 & 1 & 0 & 0 & 0 & 0 & 1
\end{pmatrix} \, .
\end{equation}
As the code is self-dual, a lowest-weight error string is $E_{[7, 1, 3]} = L_{[7, 1, 3]}$. The product code has a weight-9 error string $E_{[7, 1, 3]} \otimes E_{[7, 1, 3]}$ that is equivalent to a lowest-weight error string of the concatenated code, as shown in Fig.~\ref{fig:productcode}a. However, by numerical search, we can find that there are lower-weight errors that are not equivalent to a product of $E_{[7, 1, 3]}$ error strings. The lowest-weight errors turn out to be weight-7 errors that we will refer to as $E_{\rm prod}$. These errors consists of a single bit flip in each code block, an example of which is shown in Fig.~\ref{fig:productcode}b. In other words, $d_{\rm prod} = 7$ and therefore the code distance of the $L=2$ protocol is (at most) 7.

Subsequent concatenation steps will transform these error strings into higher-weight strings. In an $L=3$ protocol, these errors turns into $E_{\rm prod} \otimes L_{[7, 1, 3]}$ errors as shown in Fig.~\ref{fig:productcode}c. The newly introduces concatenation layers will also contain errors that are $L_{[7, 1, 3]} \otimes E_{\rm prod}$, i.e., $d$ copies of $E_{\rm prod}$ errors on $d$ different concatenation blocks, an example of which is shown in Fig.~\ref{fig:productcode}d. Both of these error strings have a weight of $d \cdot d_{\rm prod} = 21$. Each concatenation step will increase the weight of such errors by another factor of $d$.

We conjecture that these are the lowest-weight errors of blocklet protocols. If this is the case, then the distance of an $[n, k, d]^L$ protocol is $d_{\rm prod} \cdot d^{L-2} = (d_{\rm prod}/d^2) \cdot d^L$ (except for $L=1$ in which case the distance is $d$). For non-CSS-based protocol constructions, the product code will encode $XX$ and $ZZ$ fusion outcome bits in a single code, rather than having two separate codes. For the product code of the $[5, 1, 3]$ blocklet protocol, we find that $d_{\rm prod} = 5$. An example of such a weight-5 error string is shown in Fig.~\ref{fig:productcode}e. On the other hand, the product code distances of the $[4, 2, 2]$ and $[6, 4, 2]$ codes are $d_{\rm prod} = d^2 = 4$.

When we investigate the subthreshold behavior of the simulated protocols by fitting a $\sim p^A$ function (with the fit parameter $A$ and erasure rate $p$) to the points with the lowest logical error rates obtained in simulation, we find that the distance implied by the observed subthreshold scalings is in all cases above the conjectured minimum-weight error weight, as shown in Fig.~\ref{fig:subthreshold}. In some cases, the exponent of the subthreshold scaling is even significantly higher, exhibiting a much more rapid decline in logical error rate close to the threshold than implied by the code distance.

\section{Additional figures}
This appendix contains additional figures that are referenced in the main text.

\begin{figure*}[t!]
\centering
\includegraphics[width=0.75\linewidth]{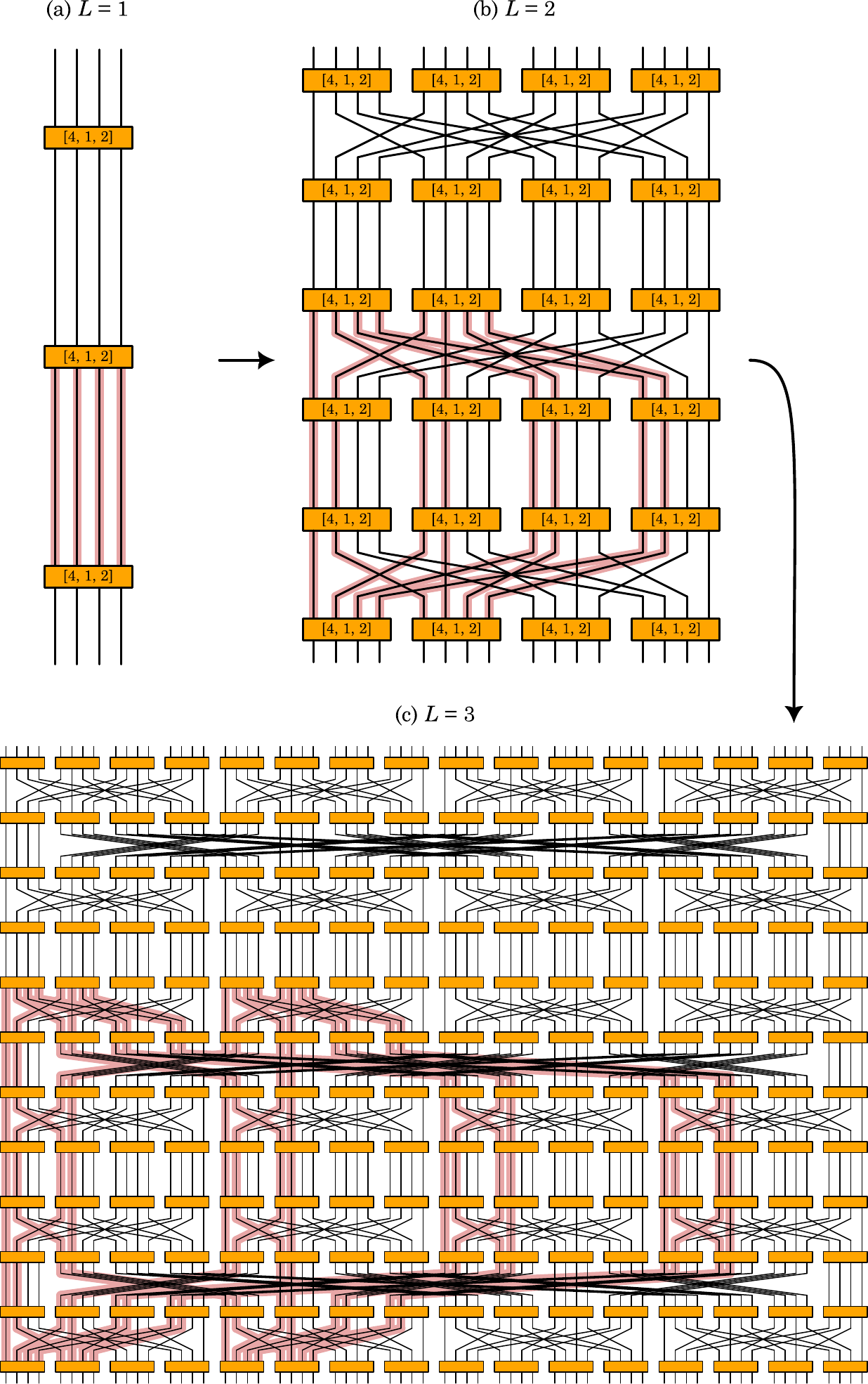}
\caption{Transformation of fundamental check through different levels of concatenation.}
\label{fig:checkconcatenation}
\end{figure*}

\begin{figure*}[t!]
\centering
\includegraphics[width=\linewidth]{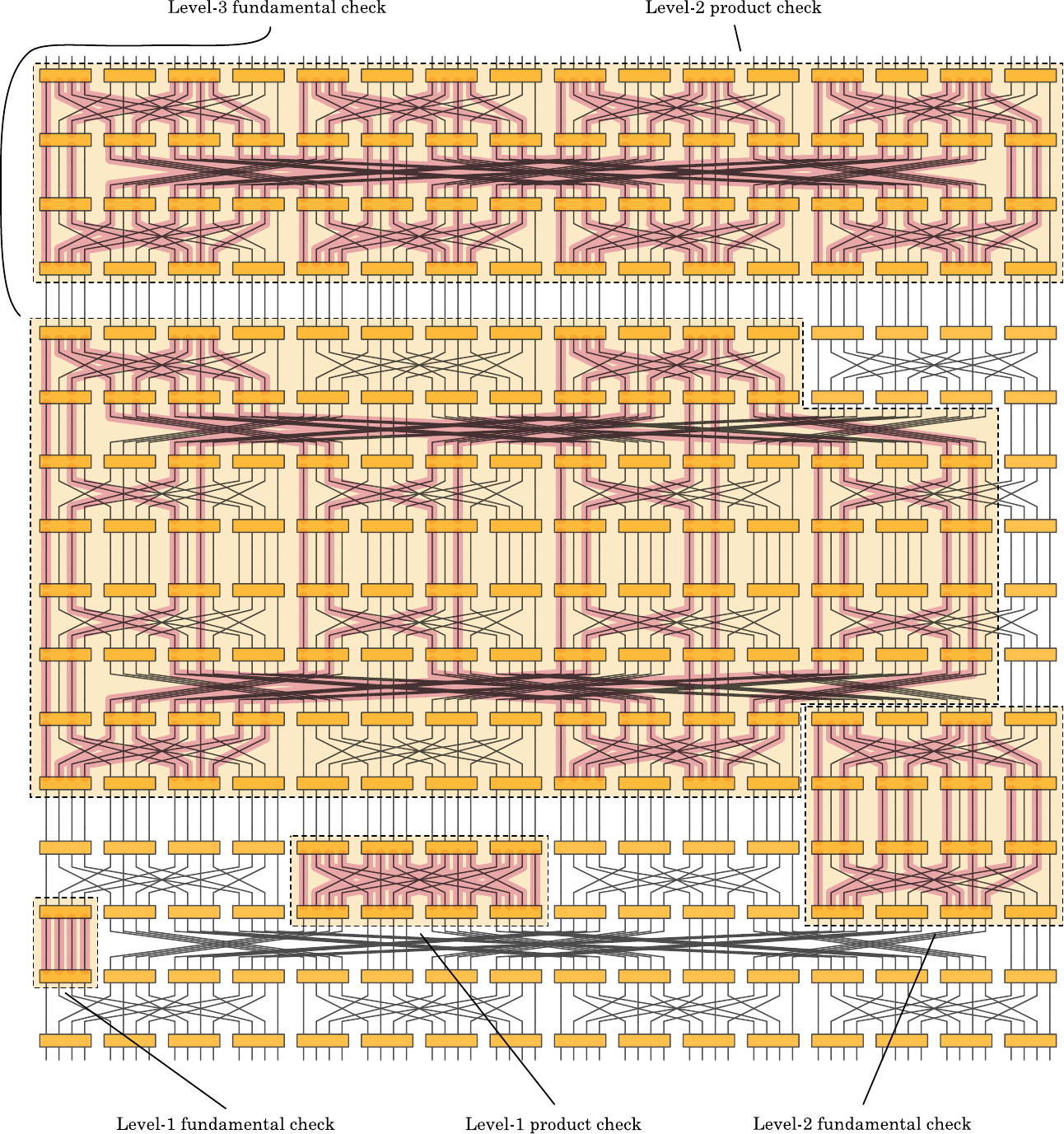}
\caption{Examples of various fundamental and product checks in an $L=3$ blocklet protocol based on a 4-qubit code.}
\label{fig:checktypes}
\end{figure*}

\begin{figure*}[t!]
\centering
\includegraphics[width=\linewidth]{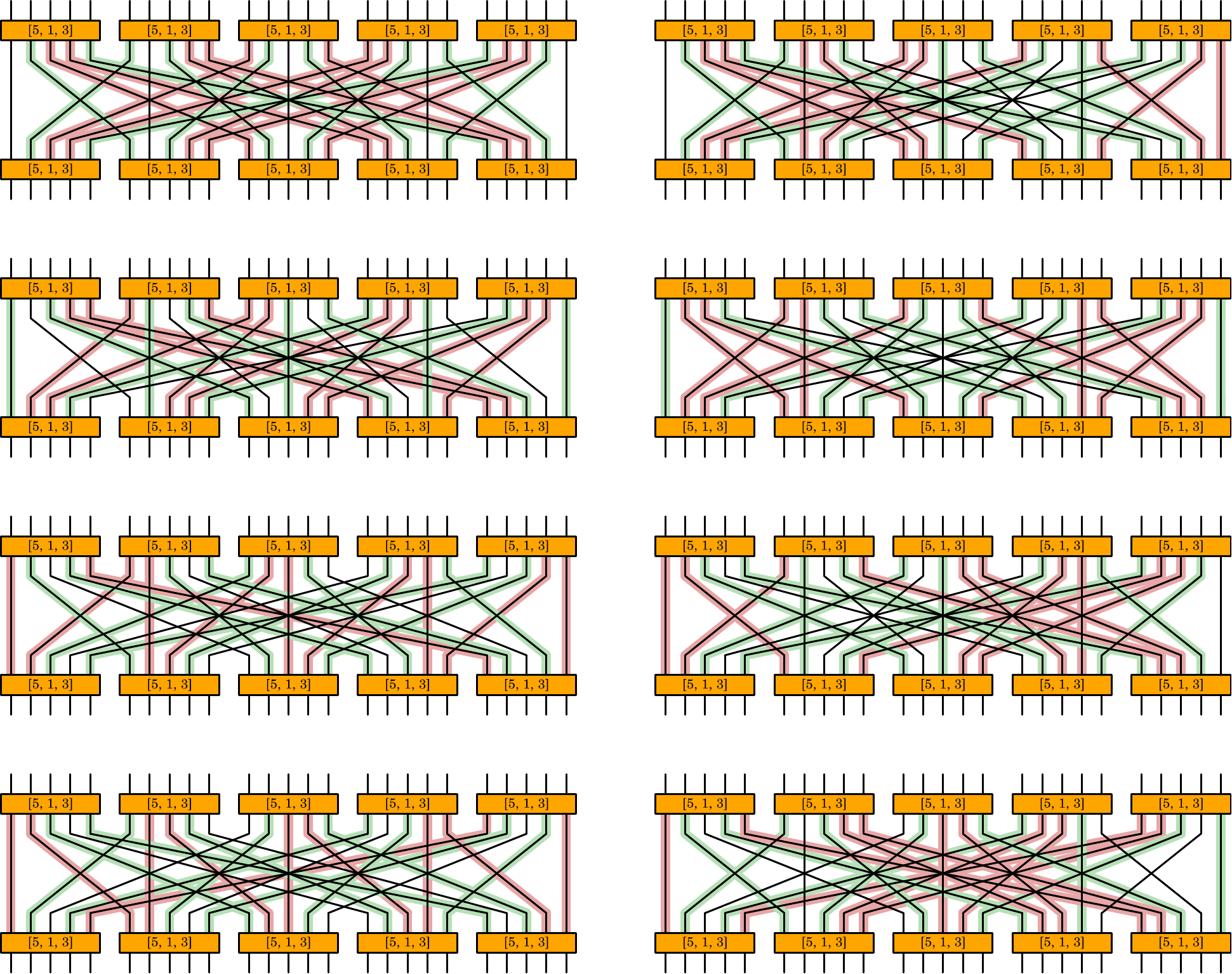}
\caption{Product checks that are generated when concatenating $[5, 1, 3]$ blocklets.}
\label{fig:513productchecks}
\end{figure*}

\begin{figure*}[t!]
\centering
\includegraphics[width=\linewidth]{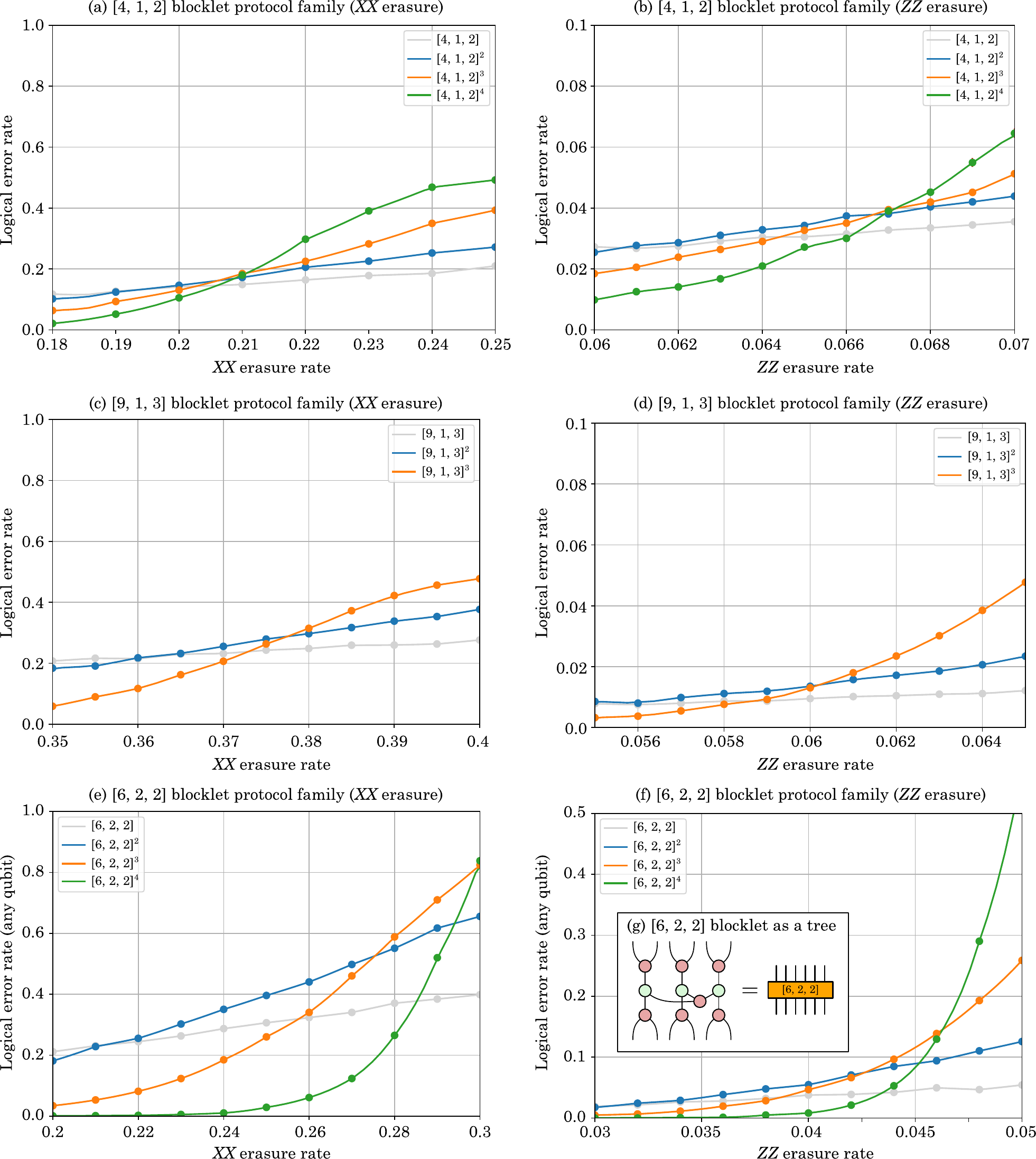}
\caption{Performance of various tree blocklet protocols.}
\label{fig:treeblocklets}
\end{figure*}

\begin{figure*}[t!]
\centering
\includegraphics[width=\linewidth]{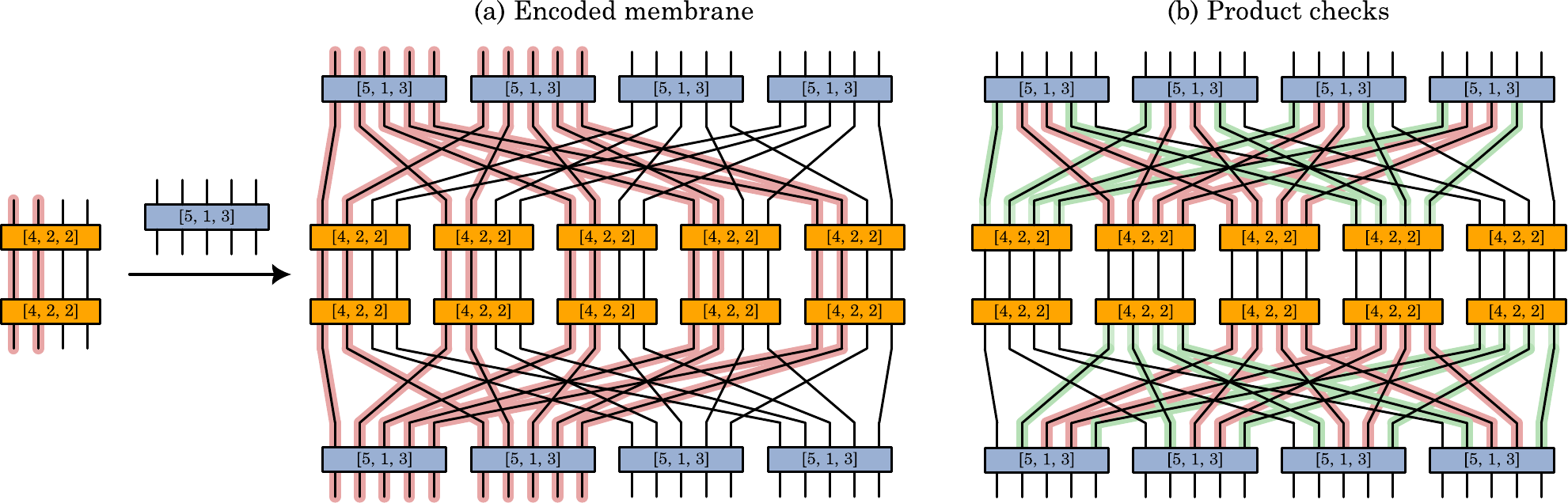}
\caption{Concatenation of inner $[4, 2, 2]$ blocklets with outer $[5, 1, 3]$ blocklets.}
\label{fig:513mixing}
\end{figure*}

\begin{figure*}[t!]
\centering
\includegraphics[width=\linewidth]{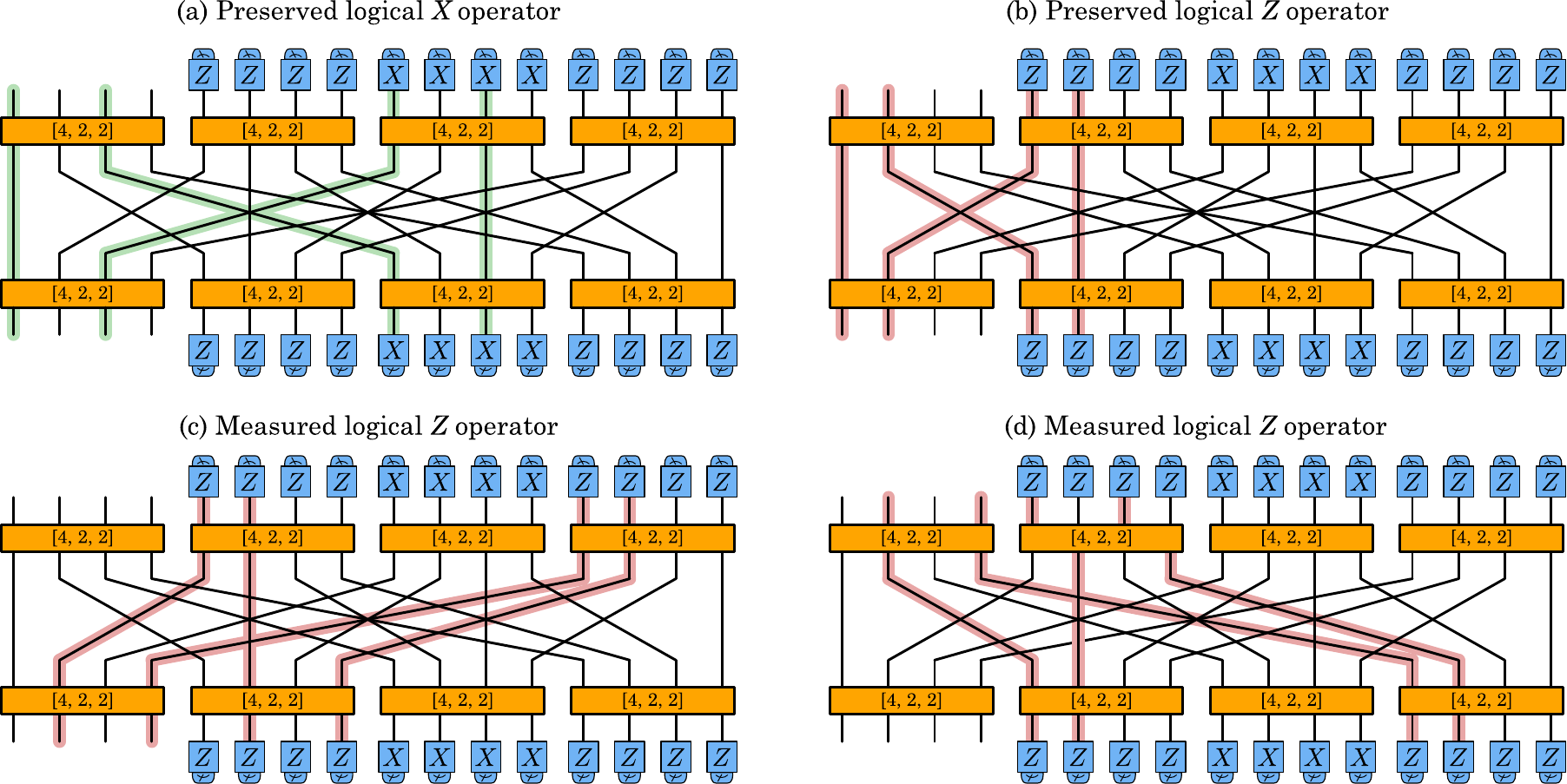}
\caption{Fault-tolerant selective measurement of one of the logical $Z$ operators in a $[4, 2, 2]$ code.}
\label{fig:selectivemeasurement}
\end{figure*}

\end{document}